\def\preprint{1}			

\PassOptionsToPackage{margin=0.81in}{geometry}

\ifdefined\preprint
  \documentclass[preprint,review,12pt]{elsarticle}
\fi
\ifdefined\wordcount
  \documentclass[final,3p,times,twocolumn]{elsarticle}
\fi
\ifdefined\final
  \documentclass[final,3p,twocolumn]{elsarticle}
\fi

\usepackage{amssymb, bm}
\usepackage{amsmath}
\usepackage{subfig}
\usepackage{caption}
\usepackage{subfig}
\usepackage{graphicx}
\usepackage{epstopdf}
\usepackage{fancyhdr}
\usepackage{geometry}
\usepackage{setspace}
\usepackage{booktabs}
\usepackage{color, soul}
\usepackage[table,x11names]{xcolor}
\usepackage{nomencl}
\usepackage{multirow}
\usepackage{setspace}
\makenomenclature

\RequirePackage{ifthen}
\renewcommand{\nomgroup}[1]{\ifthenelse{\equal{#1}{R}}{\item[\textbf{Roman letters}]}{\ifthenelse{\equal{#1}{S}}{\item[\textbf{Superscripts and  subscripts}]}{\ifthenelse{\equal{#1}{G}}{\item[\textbf{Greek letters}]}{\ifthenelse{\equal{#1}{A}}{\item[\textbf{Acronyms}]}}}}}
\biboptions{sort&compress}

\journal{Fuel}

\begin{document}

\begin{frontmatter}

\title{Interface-resolved simulation of the evaporation and combustion of a fuel droplet suspended in normal gravity}

\author{A.E. Saufi\corref{cor1}}
\ead{abdessamade.saufi@polimi.it}
\cortext[cor1]{Corresponding author}
\author{A. Frassoldati}
\author{T. Faravelli}
\author{A. Cuoci}

\address{Department of Chemistry, Materials, and Chemical Engineering “G. Natta”, Piazza Leonardo da Vinci 32, 20133 Milano, Italy}

\begin{abstract}
An interface-resolved simulation of the combustion of a fuel droplet suspended in normal gravity is presented in this work, followed by an extensive analysis on the physical aspects involved. The  modeling is based on \texttt{DropletSMOKE++}, a multiphase solver  developed for the modeling of  droplet vaporization and combustion in convective conditions. A wide range of phenomena can be described by the model, including the interface advection, the phase-change, the combustion chemistry, non-ideal thermodynamics and multicomponent mixtures. To our knowledge, this is the most detailed simulation performed on this configuration, providing a useful theoretical and numerical support for the experimental activity on this field.  A recent experimental work is used as a reference, in which a methanol droplet is suspended on a quartz fiber and ignited at different oxygen concentrations. The numerical analysis offers a detailed insight into the physics of the problem and a satisfactory agreement with the experiments in terms of diameter decay, radial temperature profiles and sensitivity to the oxygen concentration. The vaporization rate is affected by the thermal conduction from the fiber, due to the high temperatures involved. Moreover, the fiber perturbs the flame itself, providing quenching  at its surface. The combustion physics is compared to the one  predicted at zero-gravity, evidencing a lower standoff-ratio, a higher flame temperature and an intense internal circulation. The distribution of the species around the droplet shows (i) a local accumulation of  intermediate oxidation products at the fiber surface and (ii) water absorption in the liquid phase,  affecting the vaporization rate.
\end{abstract}

\begin{keyword}

droplet combustion \sep
 evaporation \sep flame \sep VOF \sep support fiber \sep methanol

\end{keyword}

\end{frontmatter}

\ifdefined \wordcount
\clearpage
\fi

\ifdefined \final
\clearpage
\fi

\printnomenclature	


\section{Introduction}
 Spray and droplet combustion technologies are adopted in a wide range of practical applications: power generation, propulsion, fuel injectors etc. The numerical simulation of burning sprays is rapidly advancing, but major difficulties remain due to the inherent complexity of the problem:  breakups, coalescence phenomena, surface tension instabilities, interactions with a turbulent gas flow are very complex phenomena, not yet fully understood  \cite{tryggvason2011direct}. Phase-change and gas-phase combustion further complicate the problem, while the wide range of spatial scales involved (several orders of magnitude), makes the numerical modeling extremely expensive in terms of computational resources.  Hence, the need of simplified but still physically representative configurations: in this context, the study of the evaporation and combustion of a single isolated droplet represents an essential step towards the better understanding of  a complex system like a spray. \par 
 The research on isolated droplets has been of interest for nearly 70 years, as testified by the numerous theoretical and experimental works in the field \cite{faeth1979current, law1982recent, helenbrook2002quasi, okajima1975further, dietrich1996droplet, cuoci2005autoignition, farouk2014isolated, cuoci2017flame, morin2004vaporization, birouk2000turbulence, chauveau2019analysis, avedisian1988low, yang1991combustion, jackson1991soot, liu2016comprehensive, xu2017combustion}. These works are typically based on relatively large droplets ($0.2-1$ mm) if compared to the ones generated by a spray atomizer ($10-100$ $\mu$m). This aspect is often adopted to question on the relevance of single droplet investigations for practical applications. It is worth pointing  out the importance of studying isolated droplets, which is very relevant for different reasons: 
 
 \begin{itemize}
 	\item It allows to isolate and focus on the phase-change and reactive processes, decoupling them  from the interactions typically involved in a spray such as breakups, coalescence and  fragmentation. It also serves as a simplified/ideal system to investigate the sensitivity and quality of the  sub-models included in the numerical description (e.g. kinetic mechanism, species mass diffusion etc.);
 	\item It provides a basis for refining the understanding of vaporization, ignition and extinction processes and allows to quantitatively assess the capability to predict these  phenomena and their mutual interaction (e.g. the diffusion-reaction interaction, the effect of buoyancy on evaporation, the role of radiation on flame extinction, the effect of convection on the ignition properties etc.);
 	\item Sprays are commonly treated as a collection of zero-dimensional points tracked in a Lagrangian way, adopting a RANS/LES approach for turbulence modeling. The coupling between the flow and the particles requires additional subgrid scale models for the droplet source terms (e.g. heat and mass transfer rates) and for the particle-flow interaction (e.g. drag force). These closure models need to be developed and validated based on the detailed analysis of simpler configurations, such as single droplets \cite{helenbrook2002quasi}.
 \end{itemize}
 
  It is clear, however, that the first step is the development of dedicated numerical simulations,  at a level of detail necessary to describe of the physico-chemical phenomena involved in droplet combustion. In the last decades, researchers mainly focused on a simple, idealized condition in which a spherical droplet was investigated in microgravity \cite{okajima1975further, dietrich1996droplet}, because of the absence of buoyancy effects and droplet deformation. This was also attractive for the relatively simple mathematical modeling, exploiting the spherical symmetry of the system and leaving room for a more detailed chemistry description \cite{cuoci2005autoignition, farouk2014isolated, cuoci2017flame}. However, even if the microgravity condition still represents a valuable method for droplets analysis, there is a strong interest in considering more realistic situations in which convection and gravity play a significant role (like in sprays).  The most common case is represented by the vaporization, ignition and combustion of a single  droplet suspended in normal gravity, for which a very large number of experimental works is available  \cite{morin2004vaporization, birouk2000turbulence, chauveau2019analysis, pfeil2013effects, chauveau2000effects, chauveau1999effects, shaw2012combustion}. Despite the amount of experiments on   convective droplet combustion, the attempts to model these systems are scarce.\par  The main reason lies in the high complexity of the mathematical description, which requires a CFD  multiphase model that incorporates: (i) a numerical method for the interface advection, (ii) the solution of a two-phase velocity field, (iii) a resolved-boundary layer model for evaluating the vaporization rate in convective conditions, (iv) a multi-region approach to model the flame-fiber thermal interaction \cite{chauveau2008experimental} and (v) the implementation of gas-phase combustion. Some noteworthy papers can be found for pure evaporation in normal gravity \cite{george2017detailed,  wang2019vaporization}, but very few numerical works on combustion are available and mainly limited to non-shrinking, motionless droplets \cite{wang2018fully} or adopting one-step/global schemes for chemistry \cite{jin2010computational}. Only recently a skeletal mechanism ($\sim$20 reactions)  was used by Ghata et al. \cite{ghata2015computational} within a multiphase approach, limiting however the application to microgravity conditions. \par This work presents a full interface-resolved numerical simulation of the vaporization and combustion of a  suspended fuel droplet in normal gravity and demonstrates the  potential of this approach in unraveling several physical phenomena not obtainable otherwise.  The recent experimental work of Yadav et al. \cite{yadav2017interferometric} has been adopted as a  reference case for the detailed analysis of convective droplet combustion: a methanol droplet  is suspended  on a quartz vertical fiber (at ambient temperature and pressure) and ignited with a spark, tracking the combustion phenomena and the droplet characteristics over time.  \par The numerical modeling is based on the \texttt{DropletSMOKE++} solver \cite{saufi2019dropletsmoke++, saufi2019experimental}, a CFD  code conceived for the vaporization and combustion modeling  of fuel droplets (including i, ii, iii, iv, v). The analysis of the numerical simulation allows to get a very detailed insight into the physics of the problem, which is the main objective of this work. The paper organization reflects this purpose, including: a description of the main mathematical model (Section 3), the numerical methodology (Section 4), the numerical modeling of the experiments  and the extended analysis on the main  physical peculiarities of the system (Section 5), such as the flame-fiber interaction, the effect of gravity on the flame properties, the main species distribution around the droplet and the impact of water absorption in the liquid phase. The paper finishes with the conclusions.

\section{Experimental setup}
The experiments are carried out in a closed combustion chamber (70x70x100 mm\textsuperscript{3}) in which a methanol droplet ($D_0=1.8$ mm) is suspended with a syringe on a quartz vertical fiber ($D_f=0.6$ mm) at 300 K and atmospheric pressure. The chamber is much larger than the maximum droplet diameter $D_0$ ($\sim$ 50$D_0$) to avoid significant changes in the boundary conditions due to oxygen depletion.  Two steel electrodes (attached to a 14.5 kV ignition transformer) initiate the combustion, which is followed in two ways: a digital camera (60 fps) tracks the droplet surface regression, while a Mach Zehnder interferometer provides the whole temperature field distribution in a non-intrusive manner. More specific details (not necessary for the present modeling work) can be found in the reference work  \cite{yadav2017interferometric}.

\section{Mathematical model}
\texttt{DropletSMOKE++} is a multiphase CFD code based on the Volume Of Fluid (VOF) methodology for the interface advection. In addition to the governing equations enforcing the conservation of momentum, energy and species mass, a detailed description of the interface thermodynamics is implemented (based on cubic Equations of State). The evaporation rate  $\dot{m}$ is directly calculated from the gas-phase boundary layer and a multiregion approach for the thermal perturbation of the fiber on the droplet is included. 

\subsection{Interface advection}
The VOF methodology \cite{hirt1981volume} is an interface capturing method within an Eulerian formulation, adopting a marker function $\alpha$ to represent the fluid phases. The marker function $\alpha$ represents the liquid volumetric fraction, assuming value 0 in the gas-phase and value 1 in the liquid phase. The following equation for $\alpha$  is solved:

\begin{equation}
\frac{\partial \alpha}{\partial t} + \nabla\cdot\left(\textbf{v}\alpha\right)=\frac{\dot{m}}{\rho}-\frac{\alpha}{\rho}\frac{D\rho}{Dt}
\label{alphaEquation}
\end{equation}

where the source terms represent the contribution of the evaporation/condensation rate (depending on the sign of $\dot{m}$) and the droplet dilation due to the change of density $\frac{D\rho}{Dt}$. The interface is advected using the \texttt{isoAdvector} library \cite{roenby2016computational} developed by Roenby and Jasak, reconstructing the interface with a geometrical methodology. As shown in the reference paper, the \texttt{isoAdvector} is based on the usage of efficient isosurface calculations to estimate the distribution of fluids inside computational cells and its performances are superior to the semi-empirical compressive scheme commonly adopted in $\texttt{OpenFOAM}^{\textregistered}$ \cite{greenshields2015openfoam} in terms of shape preservation, volume conservation, interface sharpness and efficiency.

\subsection{Governing equations}
The velocity field is shared between the two phases, solving a single Navier-Stokes equation in the whole computational domain:

\begin{equation}
\frac{\partial\left(\rho\textbf{v}\right)}{\partial t} + \nabla\cdot\left(\rho\textbf{v}\otimes\textbf{v}\right)= \nabla\cdot\mu\left(\nabla\textbf{v}+\nabla\textbf{v}^T\right)-\nabla p_{rgh} -\textbf{g}\cdot\textbf{x}\nabla\rho
\end{equation}

where $p_{rgh}=p-\textbf{g}\cdot\textbf{x}$ is the dynamic pressure, which greatly simplifies the definition of boundary conditions. The momentum equation is coupled with the following continuity equation:

\begin{equation}
		\frac{1}{\rho}\frac{D\rho}{Dt}+\nabla\cdot \textbf{v} = \dot{m}\left(\frac{1}{\rho_L}-\frac{1}{\rho_G}\right)
		\label{continuityEquation}
\end{equation}

in which the $\dot{m}\left(\frac{1}{\rho_L}-\frac{1}{\rho_G}\right)$ term provides the interfacial velocity jump due to the phase-change (i.e. Stefan flow). Additionally, the energy equation is included:

\begin{equation}
\rho C_p\left(\frac{\partial T }{\partial t}+\textbf{v}\cdot\nabla T\right) = \nabla\cdot\left(k\nabla T\right) - \nabla\cdot\textbf{q}_{rad} -\nabla T\cdot\sum_{i}^{Ns}\textbf{j}_i C_{p,i} - \sum_{i}^{Ns_{L}}\dot{m}_i\Delta h_{ev,i}-\sum_{j}^{N_R}R_j\Delta H_{R,j}
\label{Teqn}	
\end{equation}

 The $\sum_{i}^{Ns_{L}}\dot{m}_i\Delta h_{ev,i}$ term accounts for the interface cooling due to the evaporation of the $Ns_{L}$ liquid species, while the $\sum_{j}^{N_R}R_j\Delta H_{R,j}$ term represents the energy source due to the $N_R$ combustion reactions. $\textbf{q}_{rad}$ describes the radiative heat transfer contribution. $\textbf{j}_i$ are the diffusion fluxes based on the species mole fractions gradient $\nabla x_i$ \cite{bird2002transport}:
 
 \begin{equation}
 \textbf{j}_i=-\rho\mathcal{D}_i\frac{M_{w,i}}{M_w}\nabla x_i
 \label{diffusionFlux}
 \end{equation}

 The transport equations for the species in the gas phase are:

\begin{equation}
\frac{\partial \rho \omega_i^G }{\partial t}+\nabla\cdot\left(\rho\textbf{v}\omega_i^G\right)  = -\nabla\cdot\textbf{j}^G_i+\sum_{j}^{N_R}R_j\nu_{i,j}
\label{speciesequation} 
\end{equation}

 where the species source term $\sum_{j}^{N_R}R_j\nu_{i,j}$ is due to the $N_R$ combustion reactions. $\nu_{i,j}$ represent the stoichiometric coefficient of the $i$-th species in the $j$-th reaction. Finally, the liquid species transport is needed for multicomponent liquids. The only difference is that a source term $\dot{m}_i$ has to be explicitly included  to account for the amount of species $i$ lost at the liquid interface for evaporation ($\dot{m}_i<0$) or added by condensation ($\dot{m}_i>0$):
 
 \begin{equation}
 \frac{\partial \rho \omega_i^L }{\partial t}+\nabla\cdot\left(\rho\textbf{v}\omega_i^L\right)  = -\nabla\cdot\textbf{j}^L_i+\dot{m}_i
 \label{speciesequationMulticomponent} 
 \end{equation}

Liquid-phase reactions are not considered. Equation \ref{speciesequationMulticomponent} will be adopted when dealing with water condensation on the surface (Section 5.7).

\subsection{Interface thermodynamics}

At the interface, vapor-liquid equilibrium conditions are assumed. Adopting a cubic Equation of State, the general equation for a two-phase systems is \cite{smithintroduction}:

\begin{equation}
p_i^0\left(T\right)x^L_i\phi_i\left(T,p_i^0\right)e^{\int_{p_i^0}^{p}\frac{v_{L,i}}{RT}dp}\gamma_i\left(T, x^L_i\right)= px^G_i\hat{\phi}_i(T,p,x^G_i)
\label{equilibrium}
\end{equation}

where $p_i^0\left(T\right)$ is the vapor pressure of species $i$, $\phi_i$ is the gas-phase fugacity coefficient for the pure species and $\hat{\phi}_i$ is the gas-phase mixture fugacity coefficient. The exponential term represents the  Poynting correction, while $\gamma_i$ is the activity coefficient for non-ideal mixtures. For fuels burning at atmospheric pressure, the equation can be well approximated with a modified Raoult's law:

\begin{equation}
p_i^0\left(T\right)x^L_i\gamma_i= px^G_i
\label{equilibriumRaoult}
\end{equation}

The equilibrium gaseous mole fraction $x^G_i$ is evaluated explicitly:

\begin{equation}
x^G_i= \frac{p_i^0\left(T\right)}{p}x^L_i\gamma_i
\label{equilibriumRaoulty_i}
\end{equation}

as well as the equilibrium gaseous mass fraction $\omega_i^G$:

\begin{equation}
\omega_i^G= \frac{p_i^0\left(T\right)}{p}x^L_i\frac{M_{w,i}}{M_w}\gamma_i
\label{equilibriumRaoulty_i_omega}
\end{equation}

and assigned to the whole liquid phase. Equation \ref{speciesequation} is then solved to advect the gaseous species.

\subsection{Evaporation rate}
The vaporization flux of each liquid species $\dot{m}_i$  is directly calculated  from a species mass balance at the interface, imposing the gas and liquid fluxes (sum of diffusive and convective) to be equal at the interface:

\begin{equation}
\textbf{j}^L_i\cdot\nabla\alpha+\dot{m}\omega^L_i=\textbf{j}^G_i\cdot\nabla\alpha+\dot{m}\omega^G_i
\label{balanceInterface}
\end{equation}

where $\nabla\alpha$ applies the evaporation flux only at the interface and accounts only for the normal component of $\textbf{j}_i$. For monocomponent liquids, $\textbf{j}^L_i=\textbf{0}$ and $\omega^L_i=1$. Therefore:

\begin{equation}
	\dot{m}=\textbf{j}^G_i\cdot\nabla\alpha+\dot{m}\omega^G_i
	\label{evaporationFluxi}
\end{equation}

Equation \ref{evaporationFluxi} can be re-arranged to give the total evaporation rate $\dot{m}$ \cite{reutzsch2020consistent}:

\begin{equation}
\dot{m}= \frac{\textbf{j}^G_i}{1-\omega^G_i}\cdot\nabla\alpha
\label{evaporationFluxMonocomponent}
\end{equation}

which is equal to $\dot{m}_i$ for monocomponent fuels. If the dot product  $\textbf{j}^G_i\cdot\nabla\alpha$ is negative we have evaporation ($\dot{m}<0$), otherwise we have condensation ($\dot{m}>0$). Summing over the liquid species $N_{s,L}$ on both sides of Equation \ref{balanceInterface}, the total evaporation rate for multicomponent fuels is easily obtained:

\begin{equation}
\dot{m}= \frac{\sum_{i}^{N_{s,L}}\textbf{j}^G_i}{1-\sum_{i}^{N_{s,L}}\omega^G_i}\cdot\nabla\alpha
\label{evaporationFluxMulticomponent}
\end{equation}

The total evaporation flux $\dot{m}$ is used as a source term in Equation \ref{alphaEquation}. The evaporation flux of each species $\dot{m}_i$ (only needed for multicomponent cases) can be calculated using either sides of Equation \ref{balanceInterface} once $\dot{m}$ is known. We use the gas-phase side:

\begin{equation}
\dot{m}_i=\textbf{j}^G_i\cdot\nabla\alpha+\dot{m}\omega^G_i
\label{evaporationFluxi_i}
\end{equation}

\subsection{Droplet suspension}
In normal gravity evaporation experiments, the surface tension force $\textbf{f}_s$ suspends the liquid droplet against gravity:

\begin{equation}
	\textbf{f}_s = \sigma\kappa\delta_s\textbf{n}
\end{equation}

where $\kappa$ is the interface curvature and  $\bm{\delta}_s$ is a Dirac delta applied on the interface. Modeling surface tension is one of the main challenges in multiphase flows at  small scales (mm and below), for two main reasons:

\begin{itemize}
	\item The surface tension force is only applied at the interface and this makes its numerical discretization very difficult (i.e. the Dirac delta $\delta_s$). Standard or trivial discretization methods are not able to perfectly balance the pressure gradient and the surface tension force, developing artificial velocity spikes at the interface (called spurious currents), which can eventually  grow and break the droplet apart;
	\item Within a VOF approach the interface curvature $\kappa$ is not easily accessible, due to the discontinuous nature of the marker $\alpha$:
	
	\begin{equation}
		\kappa=\nabla\cdot\textbf{n}=\nabla\cdot\left(\frac{\nabla\alpha}{|\nabla\alpha|}\right)
	\end{equation}

	 which makes the interface normal \textbf{n} calculation very difficult. Numerical errors on $\kappa$ represents another source of spurious currents, in addition to incorrect interface discretizations.
\end{itemize}

These issue is very well known in literature \cite{brackbill1992continuum, popinet2009accurate}: available solutions either rely on simple filtering of the $\alpha$ function \cite{raeini2012modelling} or more rigorous methods both for the surface tension discretization (e.g. Ghost Fluid Method \cite{vukvcevic2017implementation}) and curvature computation (e.g. Height Functions \cite{popinet2009accurate}). While filtering techniques are proved to be non-consistent \cite{popinet2018numerical} and hardly generalizable, rigorous techniques require a great effort to be correctly implemented. In particular, the $\texttt{OpenFOAM}^{\textregistered}$ framework lacks of reliable models for surface tension driven flows and no valid and general  solution has been proposed so far. Moreover, the aforementioned methods are not proved to efficiently work in evaporative conditions: the presence of evaporation further worsens the problem, since the Stefan flow tends to destabilize the interface thickness. The research in this sense is still at the beginning and only few results are available \cite{palmore2019volume}. \par In order to overcome this problem, a centripetal force $\textbf{f}_m$ directed towards the droplet center is introduced, in order to keep the droplet attached to the fiber and suspended in the presence of a  gravity field. In this way the surface tension force is not needed anymore and it can be suppressed, eliminating parasitic currents directly from their source. The Navier-Stokes equation becomes:

\begin{equation}
\frac{\partial\left(\rho\textbf{v}\right)}{\partial t} + \nabla\cdot\left(\rho\textbf{v}\otimes\textbf{v}\right)= \nabla\cdot\mu\left(\nabla\textbf{v}+\nabla\textbf{v}^T\right)-\nabla p_{rgh} -\textbf{g}\cdot\textbf{x}\nabla\rho + \textbf{f}_m
\end{equation}

This methodology allows to model the droplet evaporation process whatever the droplet size, without worrying about the detrimental effect of spurious currents. More details about the implementation can be found in the \texttt{DropletSMOKE++} reference work \cite{saufi2019dropletsmoke++}. 

\subsection{Multiregion approach for conjugate heat transfer}
The fuel droplet is suspended on a vertical fiber and evaporated under a normal gravity field. Numerous experimental and numerical analyses \cite{yang2001discrepancies, chauveau2019analysis, liu2015effect, avedisian2000soot} showed that the tethering system can significantly affect the vaporization process from a thermal point of view.  The solid is heated by the gaseous environment and conducts heat towards the droplet, providing a preferential path for the heat flux on the liquid. This phenomenon becomes extremely important in combustion processes (due to the high gas temperature) and when adopting large fiber diameters (due to the larger surface available for the heat transfer). The  \texttt{DropletSMOKE++} code includes a multiregion approach to account for the fiber thermal perturbation, firstly presented in \cite{saufi2019experimental}. The heat transfer is modeled considering the real geometry of the fiber, with no need of semi-empirical correlations or approximate approaches to account for the tethering system. The fluid and the solid regions are independently meshed, solved  and connected with dynamic boundary conditions, providing a full detailed numerical simulation of the three-phase system. The following equation is solved for the solid phase:

\begin{equation}
\rho_s C_{p,s}\frac{\partial T_s}{\partial t} = \nabla\cdot\left(k_s\nabla T_s\right)-\nabla\cdot\textbf{q}_{rad,s}
\label{equationfiber}
\end{equation}

while the fluid temperature field is provided by Equation \ref{Teqn} and $\textbf{q}_{rad,s}$ is the radiative heat flux from the fiber. The external surface of the solid fiber is the contact boundary between the phases. The boundary conditions describe the conservation of heat fluxes across the boundary as well as the continuity of  temperature. 

\subsection{Gas, liquid and solid properties}

The fluid properties are computed with the \texttt{OpenSMOKE++} library \cite{cuoci2015opensmoke++}. In particular, the gas physical properties ($\rho^G, \mu^G, C^G_{p}, k^G, \mathcal{D}^G_i$) are based on the kinetic theory of gases, while liquid ($\rho^L, \mu^L, C^L_{p}, k^L, \mathcal{D}^L_i$, $\Delta h_{ev,i}$, $p_i^0$) and solid ($\rho_s, C_{p,s}, k_s$) properties  are evaluated based on the correlations available in the Yaws \cite{yaws2015yaws} database. The activity coefficient $\gamma_i$ for non-ideal mixtures is calculated based on the UNIFAC approach \cite{fredenslund1975group}. Within a VOF approach, the mixture properties to be used in the governing equations are computed as follows (e.g. for density $\rho$):

\begin{equation}
\rho=\rho_L\alpha + \rho_G\left(1-\alpha\right)
\end{equation}

\subsection{Combustion modeling}  
The \texttt{DropletSMOKE++} code has been extensively validated for evaporation cases against numerous experimental data \cite{saufi2019dropletsmoke++, saufi2019experimental} in a wide range of operating conditions, both in natural and forced convection. In this work, the model is extended to include the gas-phase combustion and related phenomena. In particular:

\begin{itemize}
	
	\item In order to overcome the stiffness of reacting processes, the gas-phase chemistry implementation is based on an operator-splitting approach \cite{strang1968construction}, separating transport and reaction terms within the same time step. The methodology is well established and in this work we adopt the specific implementation from Cuoci et al. \cite{cuoci2013computational} included in the \texttt{laminarSMOKE++} solver  for the modeling of laminar flames with very detailed kinetic mechanisms (hundreds of species and thousands of species). In particular, the reaction step is based on the  \texttt{OpenSMOKE++} library \cite{cuoci2015opensmoke++},  specifically developed to efficiently solve stiff ODE systems and manage detailed kinetic mechanisms.  It is important to specify that reactions only occur in the gas-phase: within a VOF approach this requires the resolution of the reactive step only if $\alpha=0$ in the computational cell of interest;
	
	\item An optically thin model is used for radiative heat transfer, considering H\textsubscript{2}O, CO, CO\textsubscript{2} and CH\textsubscript{4} as main radiating species. This is justified by the small optical depth of the system ($a_pL\vert_{max}<0.3$, where $a_p$ is the maximum local  average absorptivity and $L$ the maximum length in the domain). Planck absorption coefficients $a_{p,i}$ are calculated for each species and averaged based on the species mole fractions to obtain $a_p$:
	
	\begin{equation}
		a_{p}=\sum_{i}^{Ns_r}x^G_ia_{p,i}
	\end{equation}
	
	where $Ns_r$ are the number of radiating species. Radiation is included in the temperature equation as the divergence of the radiating flux:
	
	\begin{equation}
	\nabla\cdot\textbf{q}_{rad}=4a_p\sigma\left(T^4-T_{env}^4\right)
	\end{equation}
	
	where $\sigma$ is the Stefan-Boltzmann constant and $T_{env}=300$ K. Radiation from the solid fiber is also considered in Equation \ref{equationfiber}, with $a_{p,s}=0.93$  for quartz \cite{yaws2015yaws}:
	
		\begin{equation}
		\nabla\cdot\textbf{q}_{rad,s}=4a_{p,s}\sigma\left(T_s^4-T_{env}^4\right)
		\end{equation}
		
	$\texttt{OpenFOAM}^{\textregistered}$ also incorporates sophisticated models for radiative heat transfer, such as P1 and DOM \cite{greenshields2015openfoam}. We did not notice any significant difference for the case examined in this work ($a_pL\vert_{max}<0.3$), but, if needed, their application to more geometrically complex configurations would be immediate; 
		
	\item When a droplet evaporates in a mildly-hot (few hundreds degrees K) environment, the liquid surface reaches (after a transient period) a steady wet-bulb temperature, below the liquid boiling point $T_b$, due to the balance of the incoming heat flux and vaporization enthalpy (Equation \ref{Teqn}). The internal liquid temperature is generally lower. The evaporation rate $\dot{m}$ in these conditions  is governed by the species diffusion flux $\textbf{j}_i$ from the surface (Equations \ref{evaporationFluxMonocomponent}, \ref{evaporationFluxMulticomponent}). On the other hand, the presence of a significant thermal perturbation from the fiber (typical case in combustion processes, with tens of hundreds degrees K in the gas-phase) can push the internal liquid temperature towards the boiling point $T_b$. In this case the liquid temperature profile should reach $T_b$ and instantaneously flatten, remaining constant. \par However, Equation \ref{Teqn}  cannot predict this profile discontinuity inside the liquid phase (the cooling term $\sum_{i}^{Ns_{L}}\dot{m}_i\Delta h_{ev,i}$ is zero outside the interface) and the internal droplet temperature would continue to increase above $T_b$. In fact, the heat flux per unit volume $\dot{q}$ acting on a point in the droplet is entirely used to vaporize the liquid, maintaining the local temperature constant. Therefore, the vaporization rate in the liquid phase should be calculated as:
	
	\begin{equation}
	\dot{m} = -\frac{\dot{q}}{\Delta h_{ev}} = -\frac{\rho C_p\frac{\partial T}{\partial t}}{\Delta h_{ev}}
	\label{boilingRate}
	\end{equation} 
	
	where $\Delta h_{ev}$ is the vaporization enthalpy. Equation  \ref{boilingRate} is to be applied at every point in the liquid phase in which $T\geq T_b$, avoiding over-heating. If the liquid is a monocomponent fuel, the species vaporization flux $\dot{m}_i$ is obviously equal to $\dot{m}$. Equation \ref{boilingRate} is also valid for multicomponent droplets, with the only difference that the boiling temperature $T_b$ depends on the local composition $x^L_i$ through Raoult's law (Equation \ref{equilibriumRaoult}) and it should be calculated in every point in the liquid phase. In this case the vaporization flux $\dot{m}_i$ can be calculated from the equilibrium gas-phase mass fraction (Equation \ref{equilibriumRaoulty_i_omega}):

	\begin{equation}
	\dot{m}_i = \dot{m}\omega_i^G=\dot{m}\frac{p_i^0\left(T\right)}{p}x^L_i\frac{M_{w,i}}{M_w}\gamma_i
	\label{miCasesTbMulticomponent}
	\end{equation}

\end{itemize}
  
 The rigorous treatment of boiling is extremely difficult, both from a theoretical and numerical point of view: it requires a detailed knowledge of the nucleation phenomena for the bubble generation and an accurate representation of surface tension at small scales for the bubble dynamics, with all the consequent issues in terms of spurious currents and interface stability \cite{popinet2009accurate}. Such a detailed treatment is beyond the aim of this work, since we are mainly interested in the droplet vaporization due to the gas-phase reaction. For this purpose, the internal boiling flux contribution ($T\geq T_b$) is redistributed on the droplet interface to evaluate the interface regression (Equation \ref{alphaEquation}).\par It is important to point out that in principle boiling occurs at temperatures higher than $T_b$, due to the superheat needed for bubbles nucleation and to overcome the surface tension energy barrier. However, in the presence of heterogeneous nucleation (due to the presence of the fiber, impurities in the fuel etc.) we can safely assume the superheat to be negligible in this case. The vaporization rate calculations are summarized in Table \ref{evaporationCases} both for monocomponent and multicomponent droplets.

\begin{table}
	\centering
	
	\begin{tabular}{lll}
		\toprule
		
		 &\quad\quad\quad monocomponent $\left(i=1\right)$ & \quad\quad\quad multicomponent $\left(i>1\right)$ \\
		\midrule
		$\dot{m}$   & \quad\quad \quad
		
		$\begin{cases} \frac{\textbf{j}^G_i}{1-\omega^G_i}\cdot\nabla\alpha & \mbox{if } T<T_b\\ \\ -\frac{\rho C_p\frac{\partial T}{\partial t}}{\Delta h_{ev,i}} & \mbox{if } T\geq T_b
		\end{cases}$
		  & \quad\quad\quad 		$\begin{cases} \frac{\sum_{i}^{N_{s,L}}\textbf{j}^G_i}{1-\sum_{i}^{N_{s,L}}\omega^G_i}\cdot\nabla\alpha & \mbox{if } T<T_b\\ \\ -\frac{\rho C_p\frac{\partial T}{\partial t}}{\Delta h_{ev, mix}} & \mbox{if } T\geq T_b
		  \end{cases}$  \\
		\midrule
		$\dot{m}_i$   &\quad\quad\quad   $\dot{m}$ & \quad\quad\quad  $\begin{cases} \textbf{j}^G_i\cdot\nabla\alpha+\dot{m}\omega^G_i & \mbox{if } T<T_b\\ \\ \dot{m}\frac{p_i^0\left(T\right)}{p}x^L_i\frac{M_{w,i}}{M_w}\gamma_i & \mbox{if } T\geq T_b
		\end{cases}  $\\
		
		\bottomrule
		
	\end{tabular}
	\caption{Summary of the vaporization rates $\dot{m}$ (total) and $\dot{m}_i$ (for each species) for monocomponent and multicomponent liquids.}
	\label{evaporationCases}
\end{table}

  \section{Numerical setup}
  The \texttt{DropletSMOKE++} solver is based on the $\texttt{OpenFOAM}^{\textregistered}$ framework,  adopting a finite-volume discretization of the governing equations. These are solved in a segregated approach and adopting the PIMPLE algorithm, a combination between SIMPLE (Semi-Implicit Method for Pressure-Linked Equations) and PISO (Pressure Implicit Splitting of Operators), to manage the pressure-velocity coupling \cite{greenshields2015openfoam}. The time step size $\Delta t$ is controlled by the stability condition governed by the Courant number (Co$<$0.5), while time integration adopts an implicit Euler method.

\subsection{Kinetic mechanism}

  	\begin{figure}
  		\centering
  		{\includegraphics[width=.4\textwidth,height=.3\textwidth]{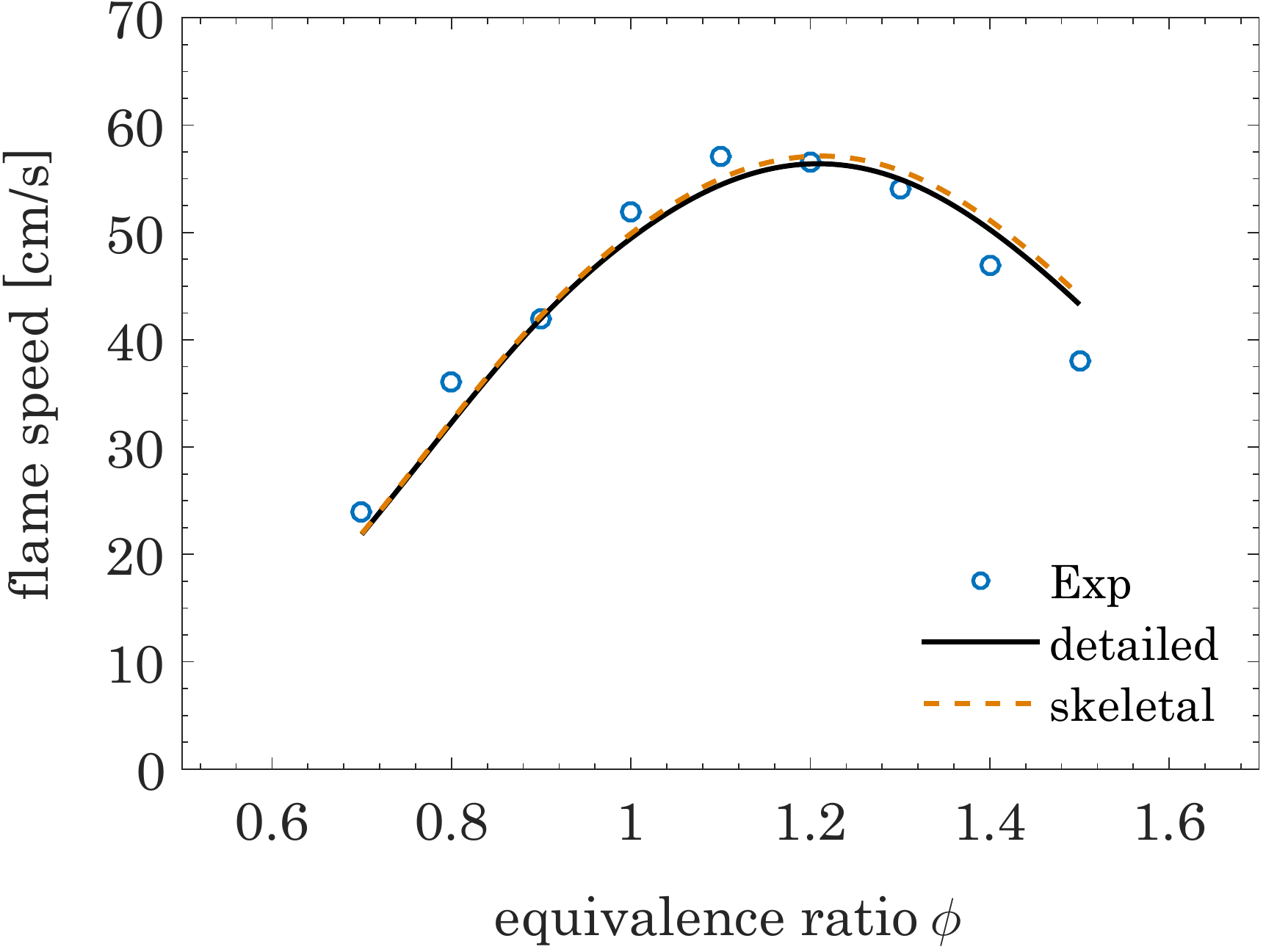}}
  		\caption{Experimental data of atmospheric laminar flame speeds of methanol at different equivalence ratios and $T_0=343$ K from Veloo et al. \cite{veloo2010comparative}. Comparison between detailed and skeletal mechanisms.}
  		\label{flameSpeedMethanol}
  	\end{figure}

The kinetic mechanism for methanol combustion was obtained from the CRECK kinetic framework \cite{ranzi2014reduced}, which describes the pyrolysis, partial oxidation and combustion of hydrocarbons up to C\textsubscript{16}.  The  C\textsubscript{0}-C\textsubscript{3} mechanism \cite{ranzi1994wide} was recently updated following the works of Metcalfe \cite{metcalfe2013hierarchical} (for H\textsubscript{2}/O\textsubscript{2} and C\textsubscript{1}/C\textsubscript{2}), Burke \cite{burke2015experimental} (for C\textsubscript{3}) and  implementing the thermodynamics from Burcat database \cite{burcat2005third}. To limit the computational cost, the resulting mechanism (115 species, 1998 reactions) was finally reduced to a skeletal mechanism (20 species, 129 reactions) using the \texttt{DoctorSMOKE++} software \cite{stagni2016skeletal}, based on a combination of the Direct Relation Graph with Error Propagation  and a species-targeted sensitivity analysis \cite{pepiot2008efficient} with a maximum error on the ignition delay time set to $8\%$. The detailed and the reduced mechanisms are compared in Figure \ref{flameSpeedMethanol}, in terms of methanol laminar flame speed. The reduced mechanism is  available in the supplementary data.

\begin{figure}
	\centering
	{\includegraphics[width=.75\textwidth,height=.6\textwidth]{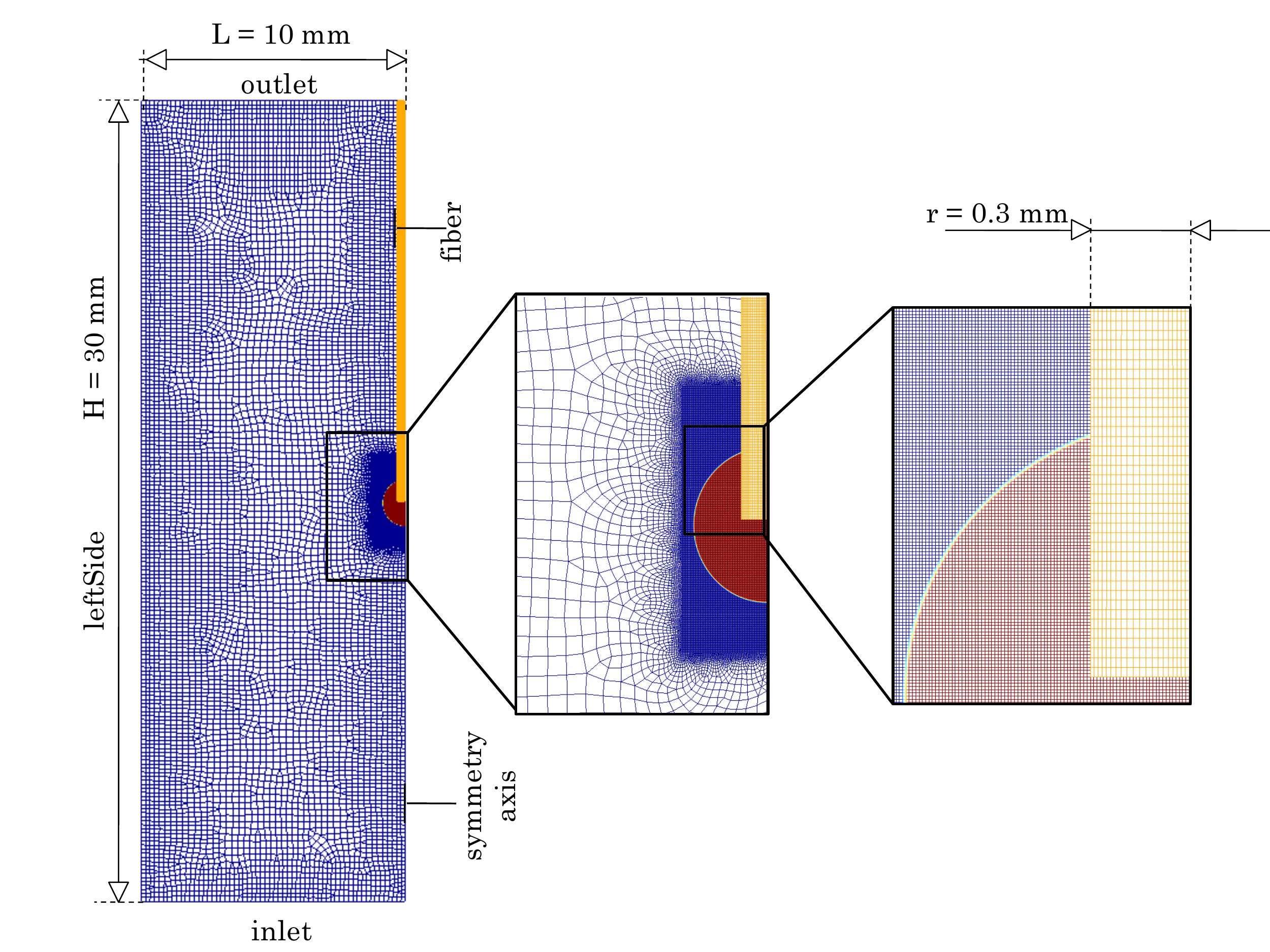}}
	
	\caption{Computational mesh used in this work, with suspended droplet (red region) at three levels of detail. The fluid region (blue) and the solid region (orange) are separately meshed and connected, sharing the \texttt{fiber} boundary.}
	\label{Mesh}
\end{figure}

\subsection{Computational mesh}
  The computational mesh is built with the commercial CFD code $\texttt{Ansys FLUENT}^{\textregistered}$  v19.2 and then converted in $\texttt{OpenFOAM}^{\textregistered}$ format. The geometry represents a cylinder  (radius $L$, height $H$) with a central vertical fiber (radius $r$) on which the droplet is suspended (Figure \ref{Mesh}). Only a slice (5 degrees) of the total geometry is modeled exploiting the axial symmetry.  The total number of cells is $\sim$ 92,000 for the fluid region and $\sim$ 10,000 for the solid region. The resulting droplet resolution is $D/\Delta x \sim 100$, necessary to solve the gas film thickness and to resolve the heat and mass transfer processes. A grid refinement analysis is reported in Appendix A, proving mesh independence. The solid and the fluid regions are meshed independently and then connected with the shared boundary $\texttt{fiber}$. The fluid region is refined around the droplet to provide a sufficiently sharp interface and resolve the boundary layer. The resulting mesh is non-structured, with a non-orthogonality coefficient equal to 57 (safe values are $<70$) and a maximum skewness of 1.56 \cite{greenshields2015openfoam}.

\subsection{Boundary conditions}  
 There are five boundaries in the computational domain named $\texttt{leftSide}$, $\texttt{inlet}$, $\texttt{outlet}$, $\texttt{symmetry axis}$ and $\texttt{fiber}$ (Figure \ref{Mesh}). The computational geometry is smaller than the real one (in order to reduce the computational cost), therefore the external boundaries $\texttt{leftSide}$, $\texttt{inlet}$ and $\texttt{outlet}$ are  considered open to not perturb the combustion process. Outlet boundary conditions are managed in $\texttt{OpenFOAM}^{\textregistered}$ as a zero gradient condition, which switches to a fixed value condition if the boundary velocity vector is directed inside the domain (backward flow). The $\texttt{fiber}$ boundary condition enforces the thermal fluxes conservation and the continuity of the temperature profile. The coupled heat transfer is included in the \texttt{turbulentTemperatureCoupledBaffleMixed} boundary condition, available in $\texttt{OpenFOAM}^{\textregistered}$ for conjugate heat transfer problems \cite{greenshields2015openfoam}. A summary of the boundary conditions is presented in Table \ref{tableboundary}. 

\begin{table}
	\centering
	
	\begin{tabular}{lllll}
		\toprule
		
		Boundary  & $\textbf{v}$ & Temperature & $\omega_i$ & $p$ \\
		\midrule
		\texttt{inlet, outlet, leftSide}   &  open & open & open & $p=p_{ext}$  \\
		\texttt{fiber}  & $\textbf{v}=0$ &  fluid-solid coupling & $\nabla \omega_i =0$ & $\nabla p =0$  \\
		\bottomrule
		
	\end{tabular}
	\caption{Boundary conditions for  velocity \textbf{v}, temperature $T$, species mass fraction $\omega_i$ and pressure $p$. The mesh is presented in Figure \ref{Mesh}.}
	\label{tableboundary}
\end{table}
  
\subsection{Parallelization}
The $\texttt{DropletSMOKE++}$ code works in parallel mode, adopting the Domain Decomposition Method already included in $\texttt{OpenFOAM}^{\textregistered}$. Almost $\sim95\%$ of the CPU time  is used for the chemical step (the resolution of the ODE systems). Since this is a local step (no data transfer across the processors is needed), the parallelization efficiency is very high. The simulations presented in this work were run on a multi-processor machine (Intel Xeon X5675, 3.07 GHz). Using 60 processors the average CPU time was $\sim90-100$ h.

\section{Combustion of methanol droplets}

\begin{table}
	\centering
	
	\begin{tabular}{lllll}
		\toprule
		Case ~~ & $D_0$ [mm] ~~& $x^G_{O_2}$ [-]    \\
		\midrule
		1     & 1.8     & 0.17 \\
		2     & 1.8     & 0.21  \\
		3     & 1.8     & 0.25 \\				
		\bottomrule
		
	\end{tabular}	
	
	\caption{Experimental cases of burning methanol droplets from Yadav et al. \cite{yadav2017interferometric} examined in this work. $x^G_{O_2}$ is the initial oxygen mole fraction in the gas-phase. $p=1$ bar, $T=300$ K.}
	\label{tableExperimentalCases}
\end{table}

\subsection{Initial conditions}

In this work three cases of methanol droplet combustion are considered, varying the oxygen mole fraction in the gas-phase. The cases are summarized in Table \ref{tableExperimentalCases}. The spark ignition is simulated as a small sphere ($D=0.1$ mm) having a temperature $T_{spark}=2500$ K placed at 1 mm from the droplet surface (on a horizontal line passing through the center). A short sensitivity analysis on the spark position did not show any substantial difference in the vaporization dynamics. The spark is  applied after a short time of pure evaporation ($\sim$ 0.02 s) to facilitate the ignition, for a total duration of 0.05 s. The initial conditions for the $\alpha$ function and for the methanol mass fraction are shown in Figure \ref{Mesh}. The fluid and solid initial temperatures are $T=300$ K and the initial pressure is $p=1$ bar.

\subsection{Ignition and droplet combustion}

\begin{figure}
	\centering
	{\includegraphics[width=.99\textwidth,height=.52\textwidth]{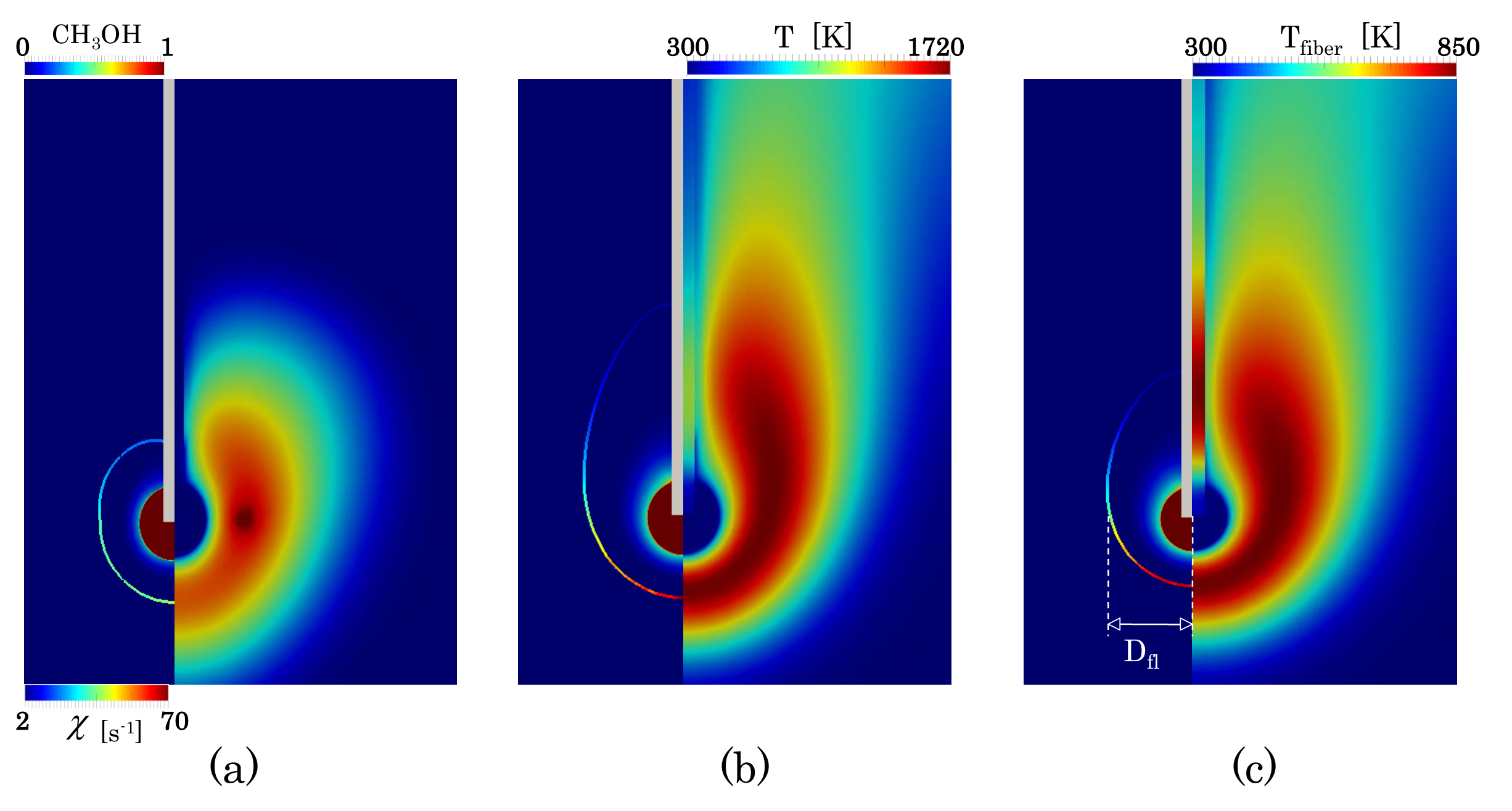}}
	
	\caption{Case 2: Maps of methanol mass fraction (left), fluid and solid temperatures (right) at times $t=0.07$ s (a), $t=0.5$ s (b) and $t=1$ s (c). On the left side the stochiometric passive scalar $Z_{st}$ contour is shown, colored by the scalar dissipation rate $\chi=2\mathcal{D}_{N_2}|\nabla Z|^2$ \cite{poinsot2005theoretical}. The flame diameter $D_{fl}$ definition is evidenced in (c).}
	\label{flameFields}
\end{figure}

\begin{figure}
	\centering
	{\includegraphics[width=.95\textwidth,height=.53\textwidth]{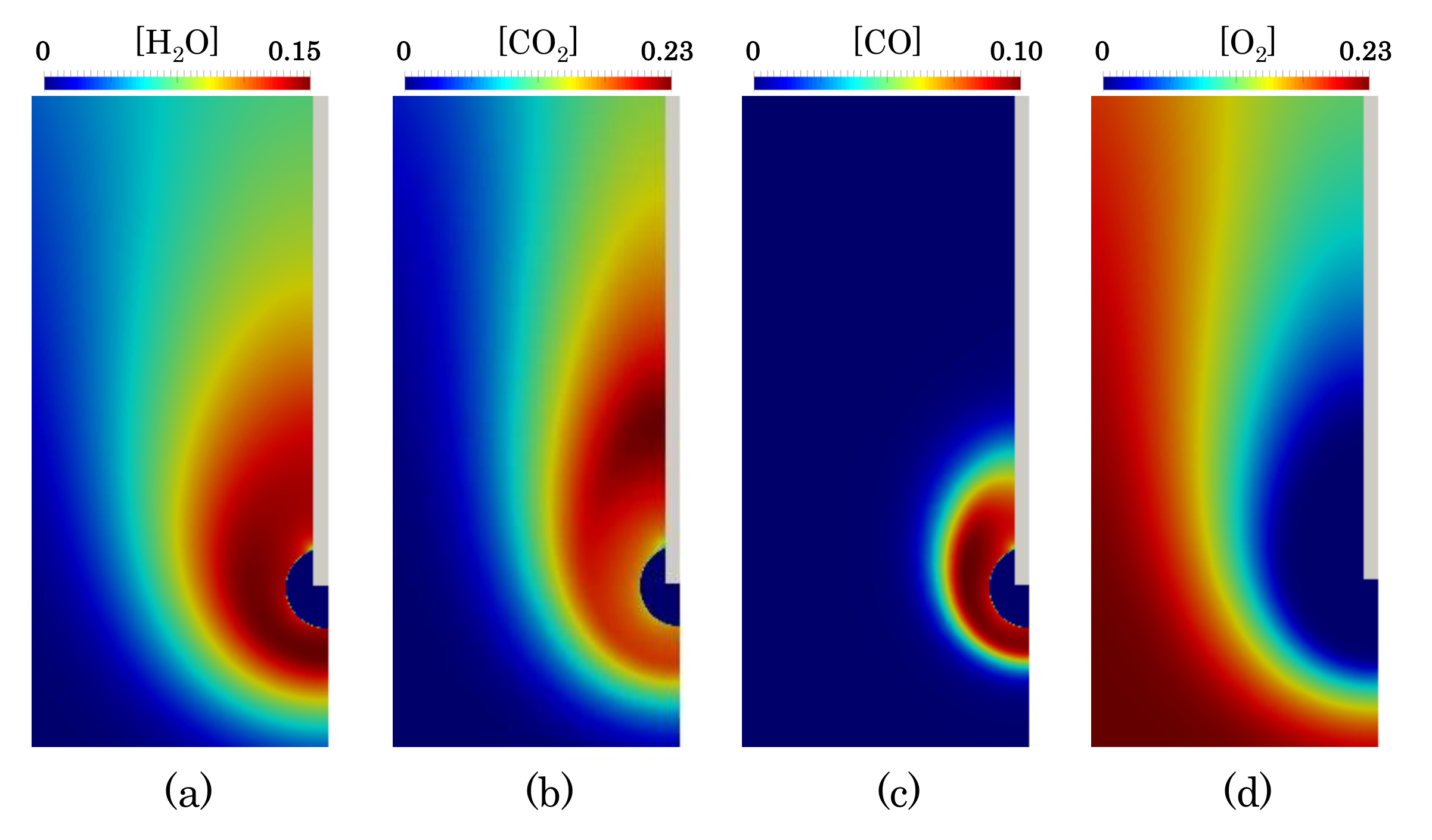}}
	
	\caption{Case 2: Maps of H\textsubscript{2}O (a), CO\textsubscript{2} (b), CO (c), O\textsubscript{2} (d) mass fractions. The gray region represents the fiber. Time $t=0.67$ s.}
	\label{flames}
\end{figure}

The ignition and combustion dynamics of Case 2 is qualitatively presented in Figure \ref{flameFields} by means of temperature and methanol mass fraction fields.  In Figure \ref{flames} the main combustion products (H\textsubscript{2}O, CO\textsubscript{2}, CO) and O\textsubscript{2} are presented in terms of mass fractions. The gas-phase ignites at $t\sim 0.07$ s, developing a buoyant diffusion flame. In Figure \ref{flameFields}a  the ignition spark is still visible close to the droplet surface. The flow regime around the droplet is  laminar ($Re<20$) and previous authors \cite{helenbrook2002quasi} report that, for spheres immersed in a convective field, values of the Reynolds number below 200 are sufficient to justify the assumption of axial symmetry. 
It is useful to analyze the flame structure adopting a passive scalar $Z$  \cite{poinsot2005theoretical}. Following Bilger's approach \cite{bilger1990reduced}, we define the parameter $\beta$:

\begin{equation}
	\beta=\frac{2}{M_{w,C}}\omega_C+\frac{1}{2M_{w,H}}\omega_H-\frac{1}{M_{w,O}}\omega_O
\end{equation}

where the mass fraction of the $k$-th element (C, H, O) are calculated as:

\begin{equation}
	\omega_k=\sum_{i}^{Ns}\omega_i N_{k,i}\frac{M_{w,k}}{M_{w,i}}
\end{equation}

where $N_{k,i}$ is the number of atoms $k$ in species $i$. The mixture fraction $Z$ is defined as:

\begin{equation}
	Z=\frac{\beta-\beta_{ox}}{\beta_{fuel}-\beta_{ox}}
	\label{mixtureFraction}
\end{equation}

The fuel composition used to calculate $\beta_{fuel}$ is the gaseous composition at the droplet interface, including the dilution with the surrounding species.  Therefore, $Z$  assumes value 1 on the interface and 0 in the pure oxidizer. The stoichiometric $Z_{st}$ is easily calculated imposing $\beta=0$ (whatever the chemical reactions involved):

\begin{equation}
Z_{st}=-\frac{\beta_{ox}}{\beta_{fuel}-\beta_{ox}}
\end{equation}

The $Z_{st}$ isocontour is evidenced in Figure \ref{flameFields}. It is  colored by the local scalar dissipation rate $\chi$:

\begin{equation}
	 \chi=2\mathcal{D}_{N_2}|\nabla Z|^2
\end{equation}

The inverse of $|\nabla Z|^2$ is proportional to the flame thickness \cite{poinsot2005theoretical}. As can be seen in Figure \ref{flameFields},  the lower part of the flame is thinner because it behaves similarly to a counterflow flame: the Stefan flow generated by the droplet vaporization  is opposed to the upward flow due to buoyancy, moving the $Z$ iso-surfaces closer and reducing the thickness of the reactive region. As a consequence, the highest concentration of radical species (OH, H, HO\textsubscript{2}) are localized in this region. The flame becomes thicker in the upper part of the flame because of the absence of a significant relative velocity component normal to the $Z$ iso-surfaces.

\begin{figure}
	\centering
	\subfloat[]		
	{\includegraphics[width=.32\textwidth,height=.25\textwidth]{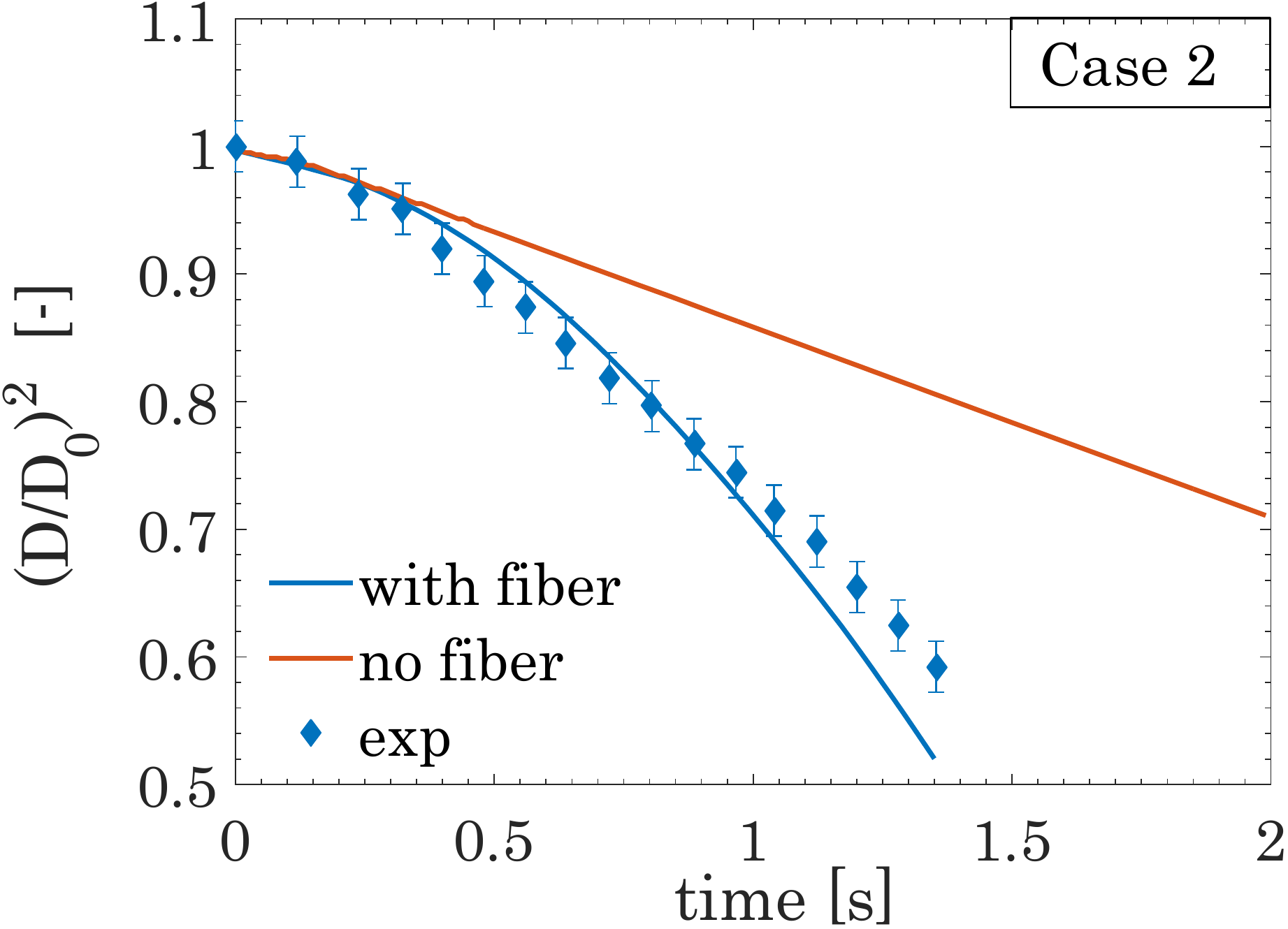}}~~
	\subfloat[]
	{\includegraphics[width=.32\textwidth,height=.25\textwidth]{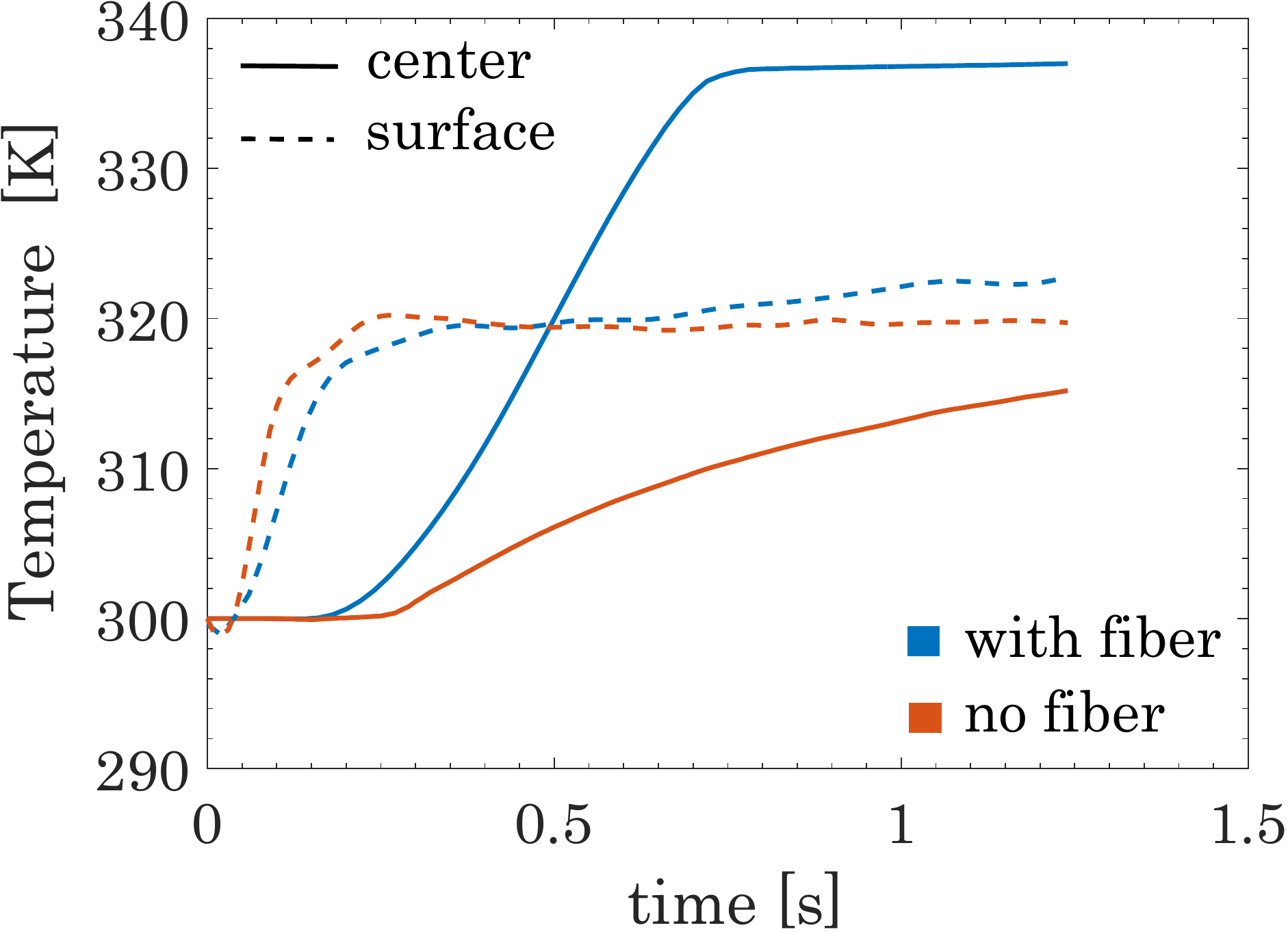}}~
	\subfloat[]
	{\includegraphics[width=.33\textwidth,height=.27\textwidth]{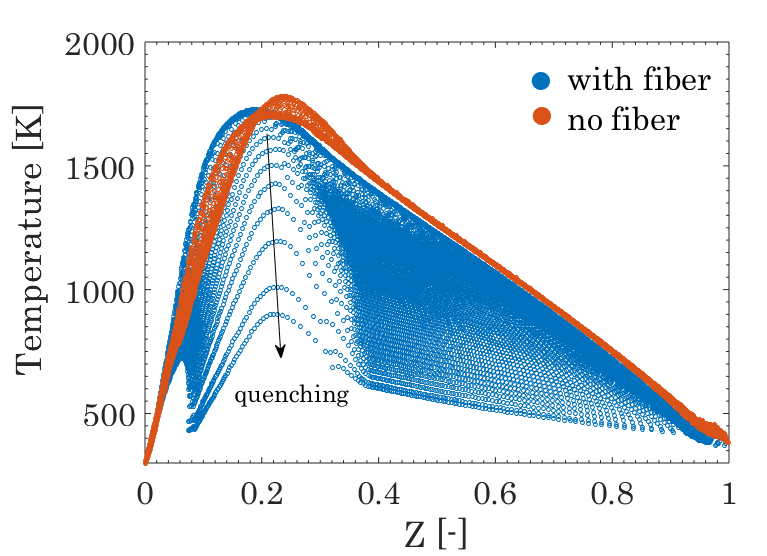}} 			
	\caption{Effect of the fiber (Case 2): $\left(D/D_0\right)^2$ plot (a),  droplet temperature (b) and $T$-$Z$ scatterplot at $t=0.67$ s  (c). Experimental data from \cite{yadav2017interferometric}.  }
	\label{profiles}
\end{figure}

\subsection{The thermal effect of the fiber}
The temperature of the fiber increases due to the vicinity of the flame (up to $\sim 850$ K,  Figure \ref{flameFields}c) and this is known to affect the combustion process \cite{chauveau2008experimental}, creating a preferential path for the heat flux on the droplet (especially for large fiber diameters). \par In order to quantify the thermal perturbation caused by the fiber, Case 2 has also been simulated considering it as adiabatic. Figure \ref{profiles}a shows that if the fiber heat transfer is neglected, the model is not able to predict the correct diameter decay, since the vaporization rate is strongly underestimated. In particular, two main processes govern the liquid temperature (Figure \ref{profiles}b): initially, after a slight decrease of $T$ due to  evaporation, the droplet surface receives heat from the flame, releasing vapor in the gas-phase and rapidly reaching an equilibrium temperature (i.e. wet-bulb temperature $\sim$ 320 K). In the meantime the fiber temperature increases, conducting heat directly inside the liquid. Since evaporation is not possible inside the droplet, the temperature rapidly increases and reaches the boiling temperature $T_b$ in the center. The evaporation rate evaluation switches to the boiling model (Table \ref{evaporationCases}) and the temperature remains constant and equal to $T_b$ (blue solid line in Figure \ref{profiles}b). This creates a situation in which the liquid surface evaporates at  $T<T_b$, while the droplet interior boils ($T=T_b$). 
At the initial stages, the heat absorbed from the droplet surface governs the vaporization while the fiber plays a major role towards the end of the simulation (when the fiber is hotter and the droplet smaller), mainly affecting the internal temperature. If the fiber (Equation \ref{equationfiber}) and the boiling (Equation \ref{boilingRate}) are accounted for, the model  predicts the diameter decay  (Figure \ref{profiles}a) with a much higher accuracy. \par Another interesting effect of the fiber is the partial quenching of the flame (Figure \ref{flameFields}) close to the fiber surface, due to its thermal inertia. In Figure \ref{profiles}c a scatterplot of the temperature with respect to the passive scalar $Z$ (Equation \ref{mixtureFraction}) is reported, using all the points in the gas-phase region ($\alpha=0$). These scatterplots are typically used for turbulent flames, but in this case they are useful to describe in a compact way the temperature profiles orthogonal to the flame ($Z_{st}$). The scatterplot clearly shows a wide quenching zone (blue region) when the fiber is accounted for: focusing on the $Z_{st} \sim 0.21$, the flame temperature, defined as the maximum temperature $T_{max}$ in the domain, decreases of $\sim1000$ K approaching the fiber along the flame front providing an incomplete combustion in this region. This effect is of primary importance since it leads to an accumulation of radicals and partial oxidation products at the fiber surface.

\begin{figure}
	\centering
	\subfloat[]
	{\includegraphics[width=.32\textwidth,height=.25\textwidth]{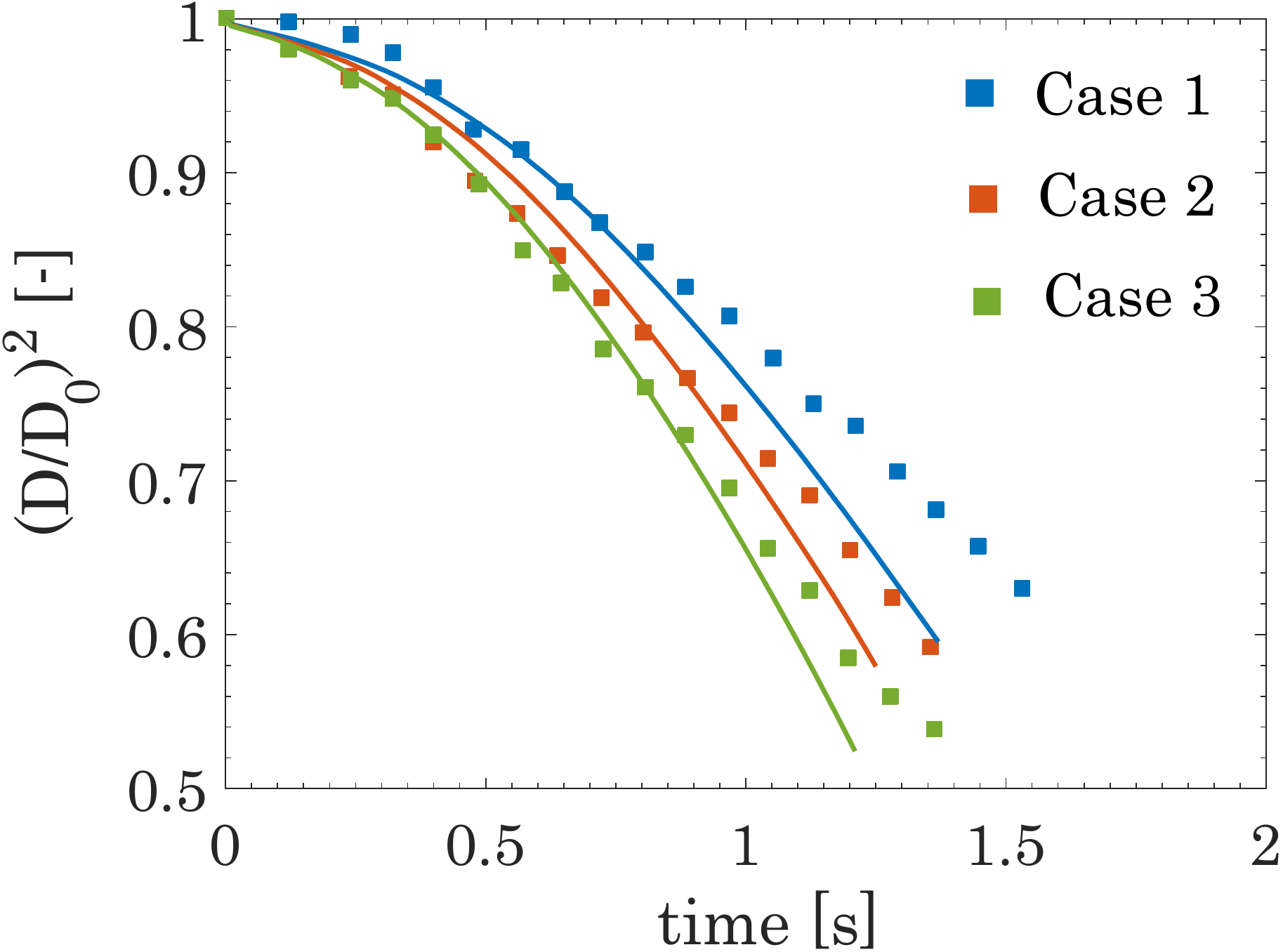}}~
	\subfloat[]
	{\includegraphics[width=.32\textwidth,height=.25\textwidth]{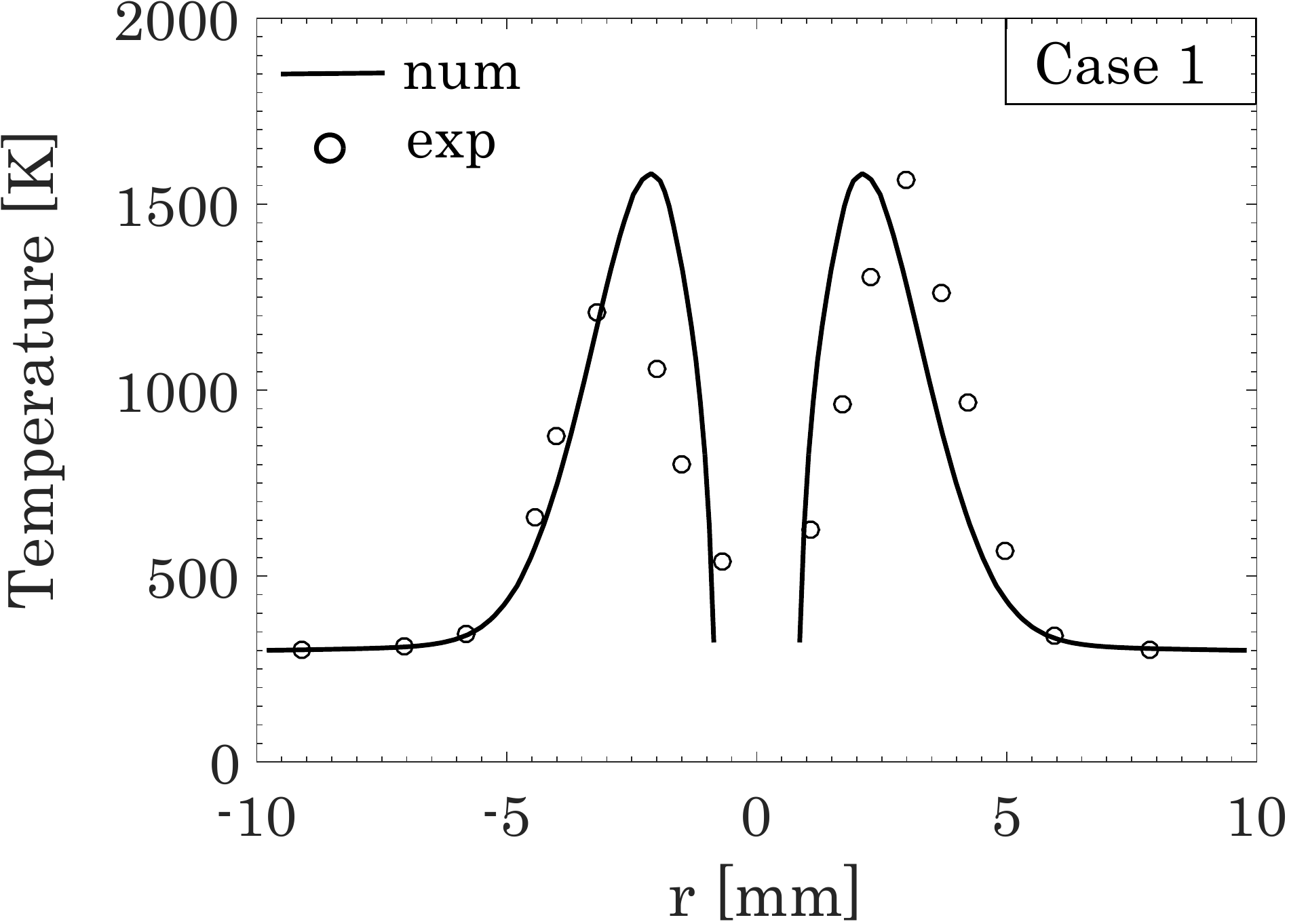}}~
	\subfloat[]
	{\includegraphics[width=.32\textwidth,height=.25\textwidth]{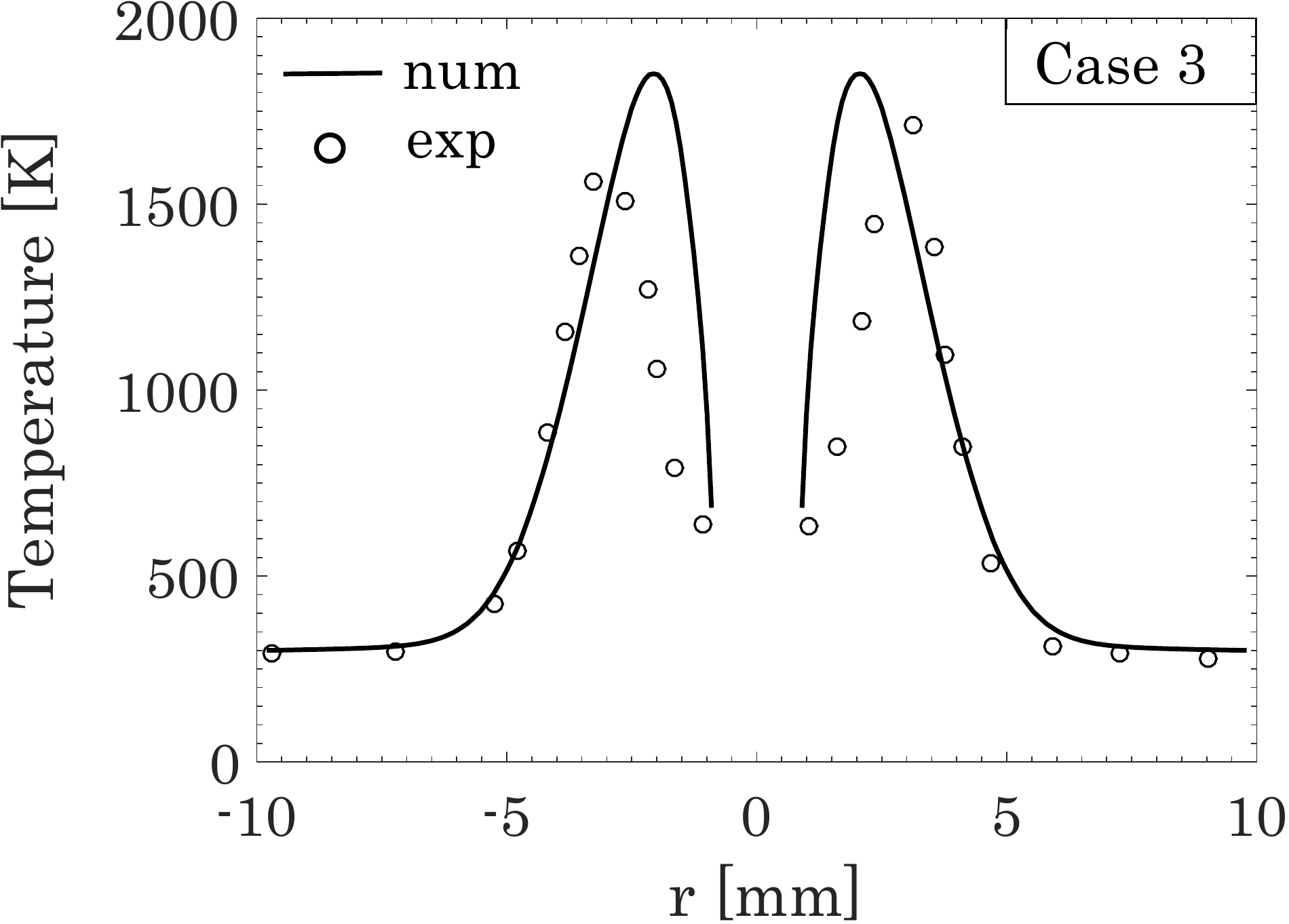}} 	 \\	
	\subfloat[]
	{\includegraphics[width=.35\textwidth,height=.25\textwidth]{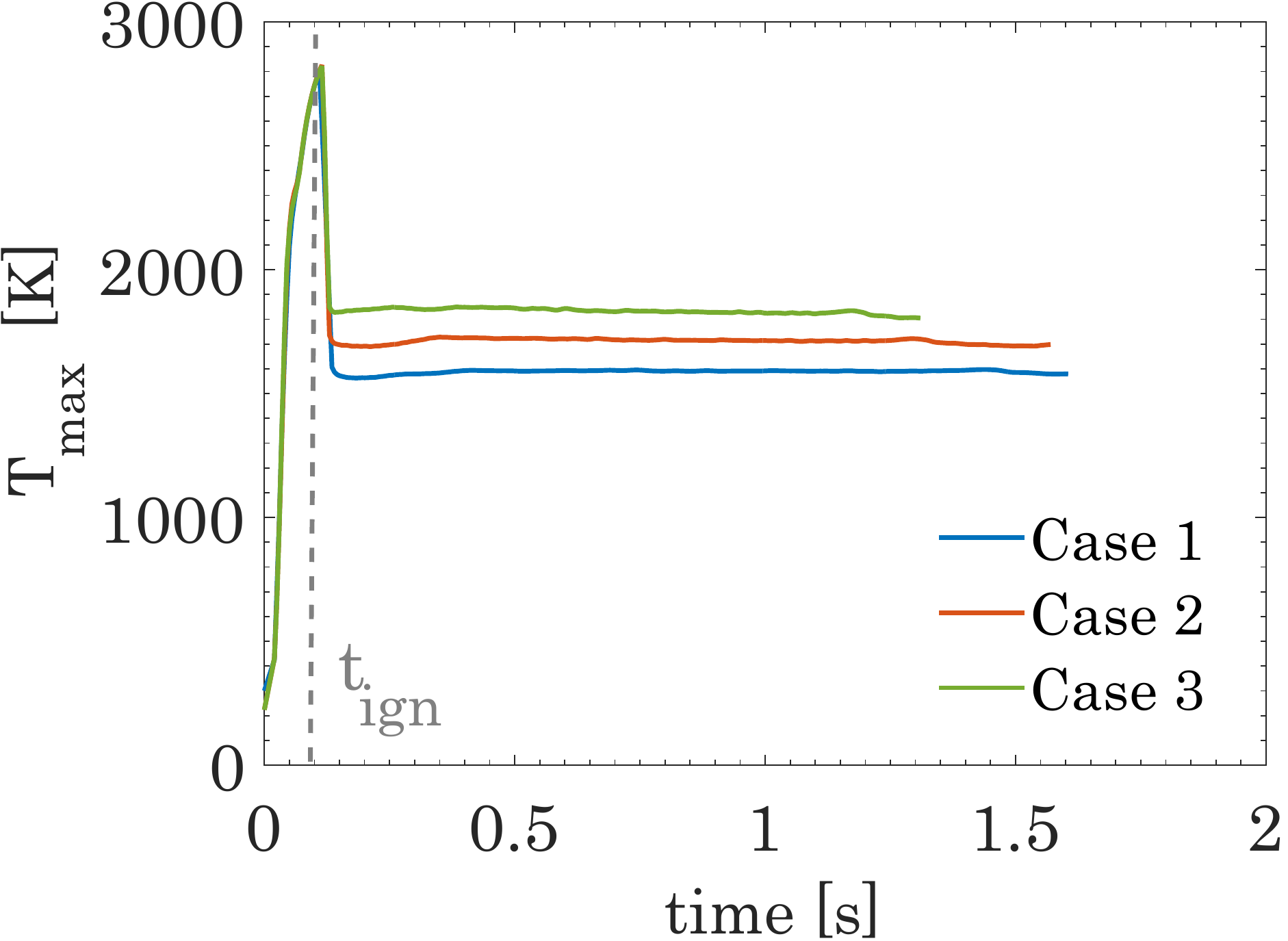}} \quad	
	\subfloat[]
	{\includegraphics[width=.33\textwidth,height=.25\textwidth]{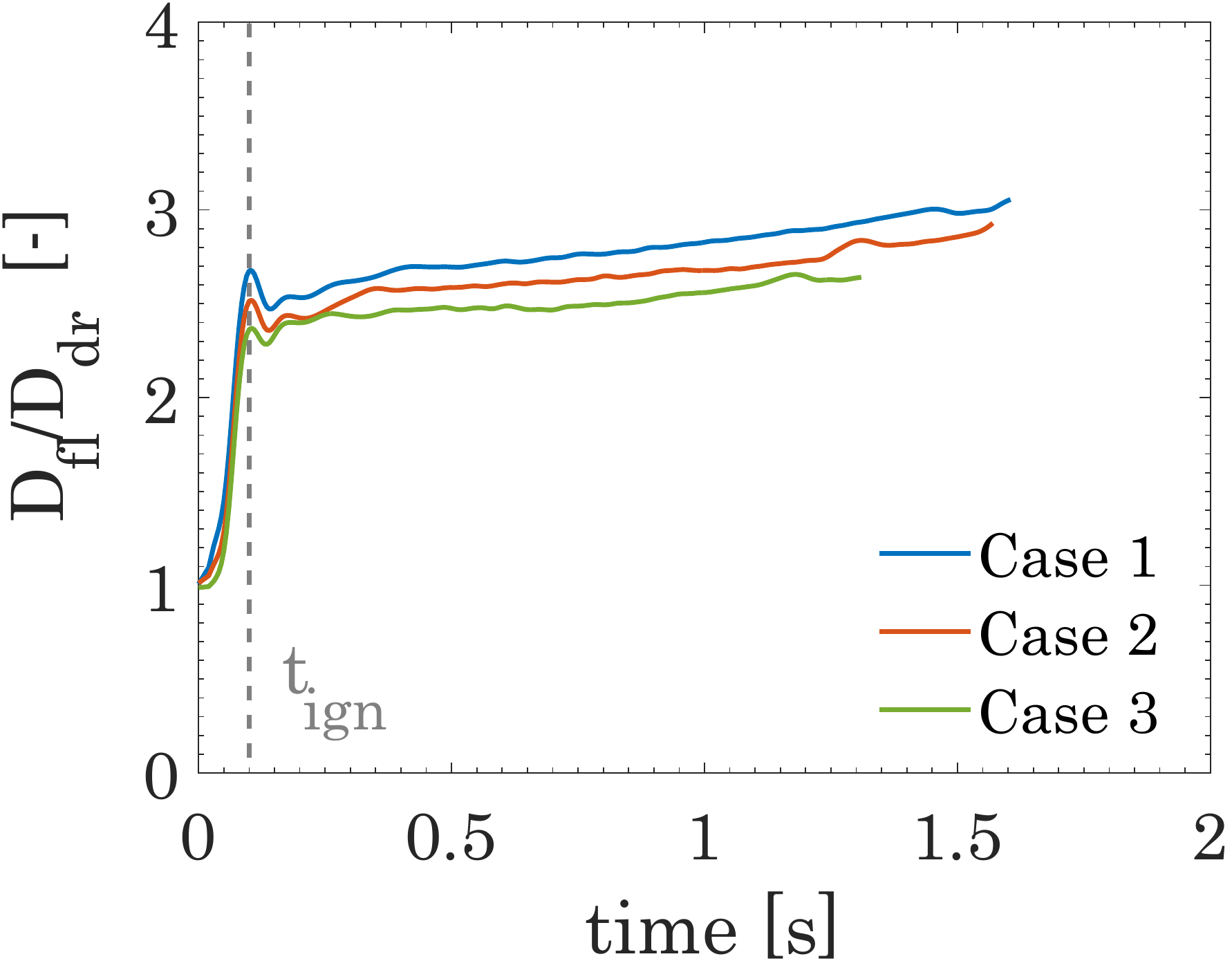}} 	 
	\caption{Effect of oxygen: $\left(D/D_0\right)^2$ plot of Cases 1, 2, 3 (a) and radial $T$ profiles of Cases 1 (b) and 3 (c) at $t=0.67$ s. Profiles of flame temperature $T_{max}$ (d) and standoff ratio $D_{fl}/D_{dr}$ (e) over time. Experimental data from \cite{yadav2017interferometric}.  }
	\label{profilesII}
\end{figure}

\subsection{The effect of the oxygen concentration}
In Figure \ref{profilesII}a the numerical diameter decays of Cases 1, 2, 3 are presented and compared with the experiments. The numerical results have been shifted, so that time $t=0$ represents the ignition time. As expected, the diameter decay is very sensitive to the atmosphere composition:  increasing the oxygen concentration leads to a higher  temperature of (i) the flame and thus (ii) of the fiber. Both induce a more intense vaporization of the liquid,  diminishing the droplet lifetime. The maximum temperature profile over time is reported in Figure \ref{profilesII}d: after the ignition the flame temperature stabilizes at a constant value, which is higher for higher oxygen concentrations in the gas-phase. This can also be observed from the radial temperature profiles, taken along an horizontal line passing through the droplet center (for Cases 1, 3 in Figures \ref{profilesII}b, \ref{profilesII}c). The model predicts the experimental data with a reasonable accuracy, considering the error associated with the measurements (up to $\sim10\%$, as reported in \cite{yadav2017interferometric}). Experimental errors can also be seen in the slight asymmetry of the temperature profiles. The difference in the maximum flame temperatures is $\sim 150$ K, with the peak for Case 3 slightly closer to the droplet surface. As reported in the last section of this paper, the agreement with the experimental results concerning the radial temperature profile improves  including water condensation on the droplet surface. 
\par Finally, it is worth analyzing the flame position with respect to the droplet interface (standoff ratio) for the three cases (Figure \ref{profilesII}e).  The flame diameter $D_{fl}$ is indicated in Figure \ref{flameFields}c and corresponds to the maximum horizontal distance of the $Z_{st}$ isoline from the symmetry axis. The flame position with respect to the droplet surface slightly increases over time for all the three cases. The flame approaches the droplet surface increasing the oxygen concentration, since the stoichiometric condition at $Z_{st}$ is satisfied closer to the droplet. The vicinity of the flame increases the vaporization rate from the droplet, further enhancing the effect of the higher oxygen concentration. This is in agreement to what observed by other authors \cite{farouk2011microgravity, farouk2012extinction}.

\subsection{The effect of gravity}
The presence of a gravity field creates an upward buoyant flow ($v_{max}\sim0.4$ m/s) once the flame is ignited and stabilized, which significantly influences the droplet combustion physics. The external convection forms a thin boundary layer, which considerably affects the heat and mass transport rates. To better discuss and highlight the effects of gravity, we simulated the cases in Table \ref{tableExperimentalCases} at zero-gravity, imposing $\textbf{g}=\textbf{0}$. The results are compared with the cases simulated in normal gravity in terms of flame temperature, standoff ratio ($D_{fl}/D_{dr}$), flame geometry and internal motion. The presence of external convection influences:

\begin{figure}
	\centering
	\subfloat[]
	{\includegraphics[width=.46\textwidth,height=.32\textwidth]{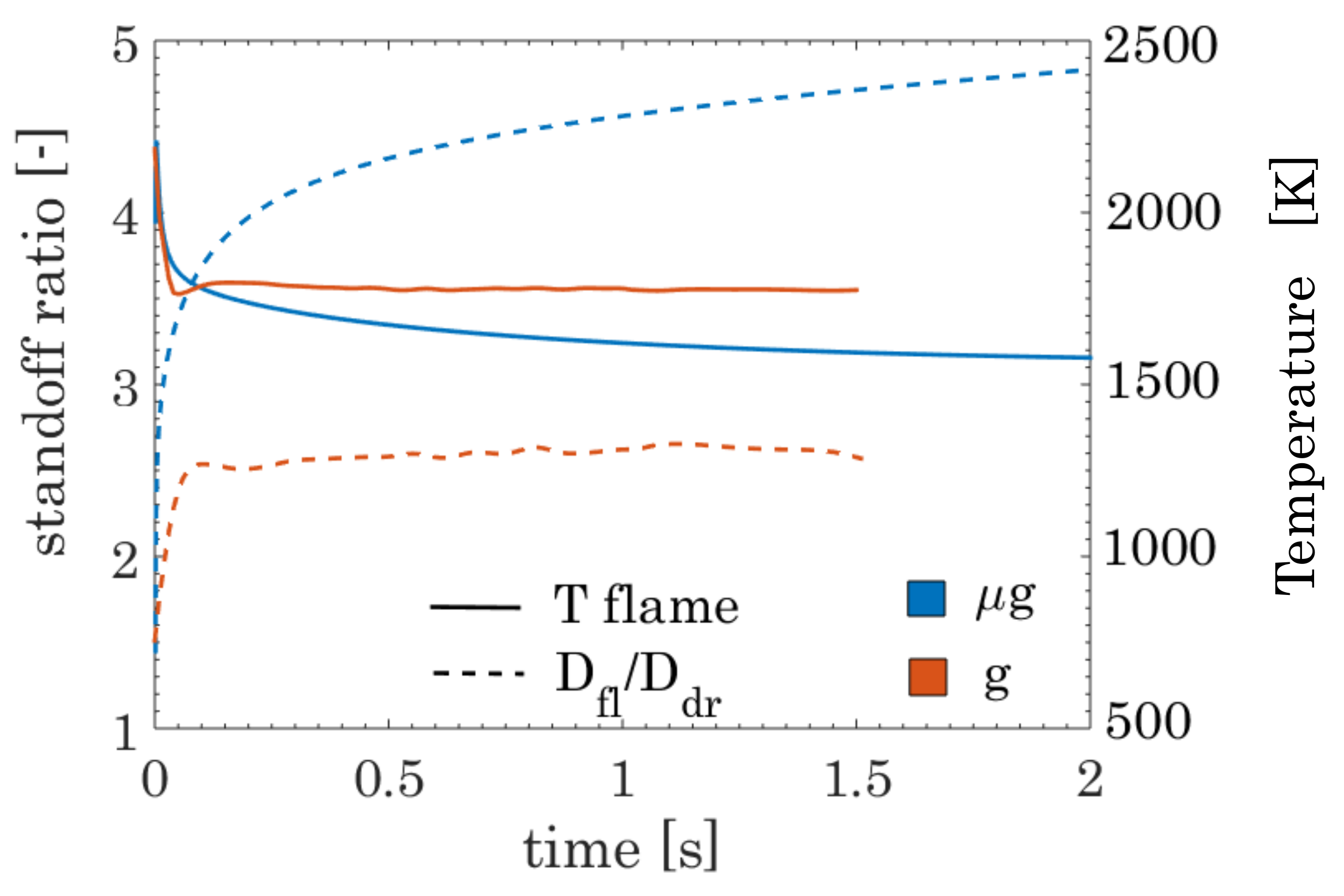}}\quad
	\subfloat[]
	{\includegraphics[width=.46\textwidth,height=.37\textwidth]{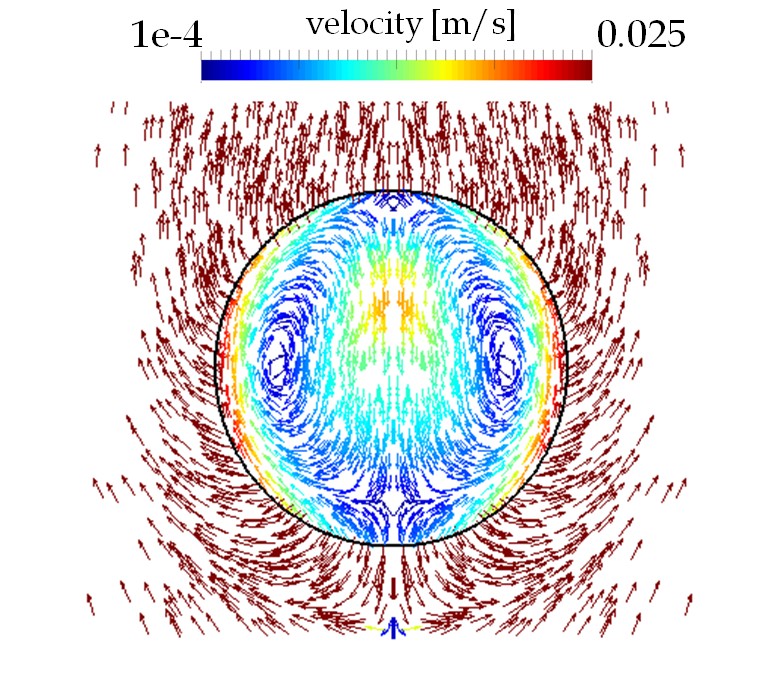}}		
	\caption{Case 2: profiles of flame $T$ and relative flame position $D_{fl}/D_{dr}$ (standoff ratio) at zero gravity and normal gravity (a). Two-phase velocity field around and inside the droplet (b), time t=1 s. }
	\label{flameStructure}
\end{figure}

\begin{itemize}
	\item \textit{The droplet shape}: under a convective field the droplet geometry is no longer spherical, due to the deformation induced by the flow. On principle this can affect the vaporization rate, since the surface area available for the vaporization is different. However, we noticed  that for small droplets this is usually a minor effect \cite{saufi2019dropletsmoke++} and that the sphericity coefficient is $\sim$1 for most of the cases of our interest;
	\item \textit{The flame geometry}: in microgravity the flame is spherical and its distance from the droplet increases after the ignition, reaching a value of $\sim$ 4-5 times the droplet diameter (Figure \ref{flameStructure}a). As a consequence the flame temperature decreases, providing a possible radiative extinction if the flame diameter is large enough \cite{cuoci2017flame}.  On the other hand, in normal gravity the flame is axisymmetric and  much closer to the droplet, maintaining its relative position almost constant in time (it is actually slightly increasing). This is due to the convective transfer of oxygen to the reactive region (much faster than pure diffusion), which satisfies the stoichiometric requirement of the flame front  at a shorter distance. The average flame thickness is significantly reduced, especially in the lower part of the flame.  As a result, the flame temperature is practically constant in time and higher than the case in microgravity (of $\sim$ 250 K).    Moreover, the flame further approaches the droplet surface when the oxygen concentration increases. We found this effect to be similar both in microgravity and normal gravity, indicating that the presence of convection does not significantly affect the response of the flame distance to the oxygen concentration;
	\item \textit{The internal circulation}: in microgravity the heat transfer in the liquid phase is mainly governed by conduction due to the absence of internal motion (if we neglect Marangoni flows). When gravity is present, the buoyancy-driven convection induces an internal motion in the liquid phase (shear stress continuity) which significantly enhances the internal heat transfer \cite{law1982recent}. Figure \ref{flameStructure}b shows the two-phase velocity field for the burning droplet, highlighting the internal motion. The internal flow structure is commonly found in suspended droplets and resembles a Hill's vortex  \cite{sirignano1999fluid}, having a toroidal core region within the droplet. The maximum velocity in the liquid phase ($\sim 2.5$ cm/s) is an order of magnitude lower than the relative gas-phase velocity ($\sim 30$ cm/s), in agreement to what observed by Prakash and Sirignano \cite{prakash1978liquid}.   \par It is worth noticing that the gas-phase velocity field exhibits a substantial radial flow at the interface, due to the Stefan flow induced by the vaporization.  Several authors observed that this radial flow leads to a significant reduction of the drag coefficient on the droplet \cite{jayawickrama2019effect}, due to the   expansion of the gas-phase boundary layer which significantly reduces the viscous force on the droplet.  As an additional effect, this can  make the liquid velocity field much less sensitive to the external convection, reducing the intensity of the internal motion. Following a fairly complex theoretical analysis, Sadhal \cite{sadhal1983flow} observes that in the limit of extremely high radial flows the internal toroidal structure  can be totally destroyed, due to the absence of a tangential component of the relative gas-phase velocity at the interface.
\end{itemize}

\subsection{Distribution of the species in the gas phase}

\begin{figure}
	\centering
	{\includegraphics[width=.95\textwidth,height=.53\textwidth]{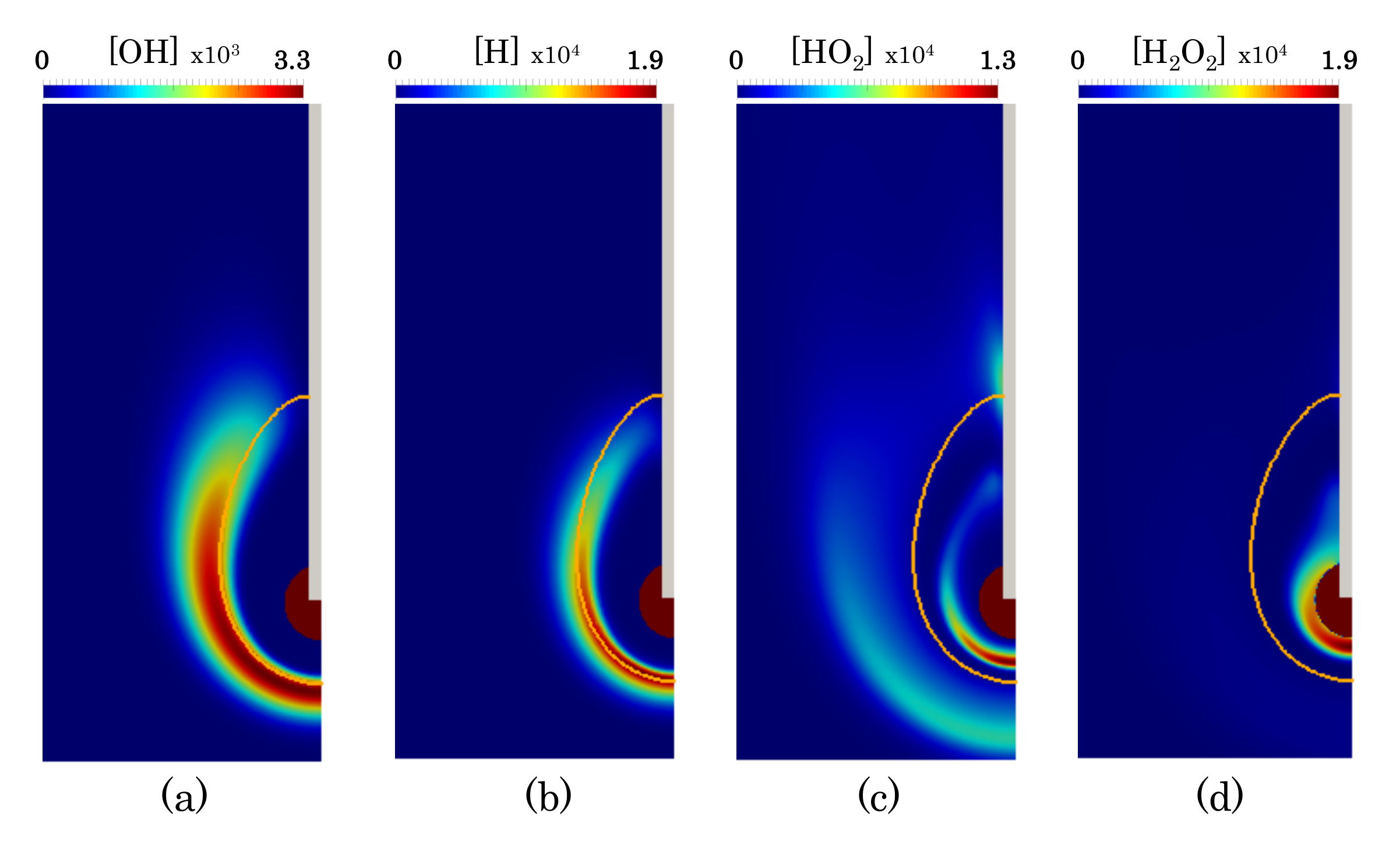}}
	
	\caption{Case 2: Maps of OH (a), H (b), HO\textsubscript{2} (c), H\textsubscript{2}O\textsubscript{2} (d) mass fractions. The orange solid line is the flame front ($Z_{st}$), the red region is the droplet. Time $t=0.67$ s.}
	\label{radicals}
\end{figure}

\begin{figure}
	\centering	
	\subfloat[]
	{\includegraphics[width=.33\textwidth,height=.27\textwidth]{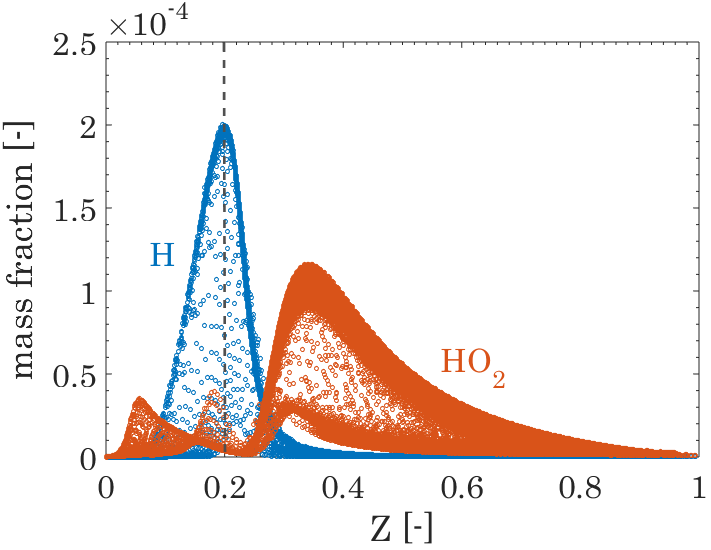}}
	\subfloat[]
	{\includegraphics[width=.33\textwidth,height=.27\textwidth]{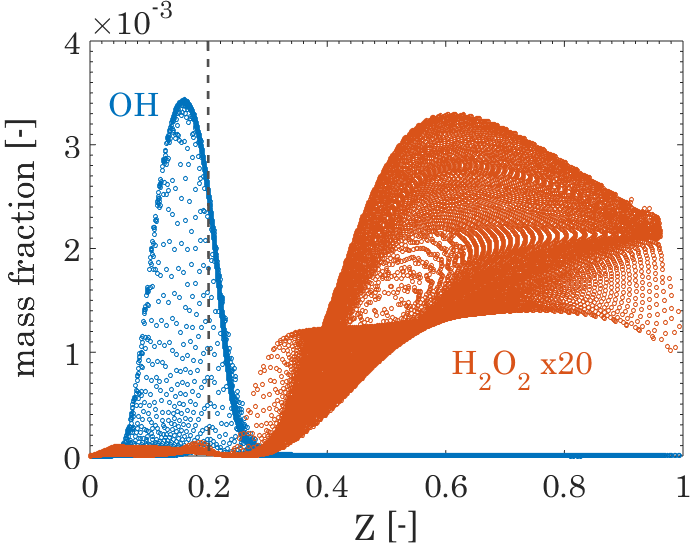}} 
	\subfloat[]
	{\includegraphics[width=.33\textwidth,height=.26\textwidth]{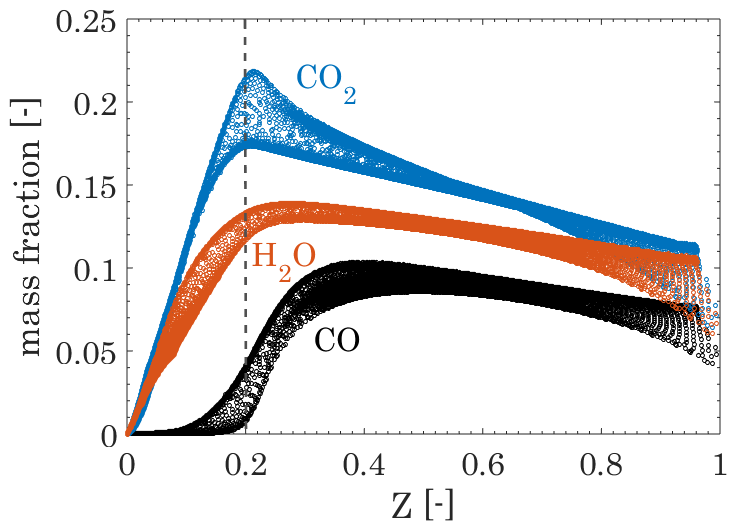}}
	\caption{Case 2: $\omega_i$-$Z$ scatterplots of main species mass fractions at time $t=0.67$ s.  }
	\label{scatterplots}
\end{figure}

Once a passive scalar $Z$ is defined (Equation \ref{mixtureFraction}), we can analyze the main species distribution (for Case 2 at $t=0.67$ s) along the flame coordinate. The results are reported in Figures \ref{radicals} and \ref{scatterplots}, highlighting the  $Z_{st}$. Analyzing the rate of production of the species, it is possible to get an insight on the methanol reactivity.  As already reported, most of the species have a peak in the high scalar dissipation rate ($\chi$) region (low thickness), the lower part of the flame (Figure  \ref{radicals}). The OH radical is slightly below the $Z_{st}$ due to the need of O\textsubscript{2} for its formation, while the H radical peak is at $Z_{st}$. The HO\textsubscript{2} radical shows an interesting profile, with three peaks along $Z$: on the rich side, HO\textsubscript{2} is mainly produced by O\textsubscript{2}+CH\textsubscript{2}OH $\rightarrow$ HO\textsubscript{2}+CH\textsubscript{2}O, while on the lean side  H+O\textsubscript{2}+(M)$\rightarrow$HO\textsubscript{2}+(M) is dominant, due to the high diffusivity of H, which escapes the  flame front. The intermediate HO\textsubscript{2} peak is due to the flame quenching at the fiber surface (clearly visible in the HO\textsubscript{2} mass fraction map in Figure \ref{radicals}c), which decreases the temperature and stabilizes HO\textsubscript{2}. The important presence of HO\textsubscript{2} for $Z>Z_{st}$ is also responsible for the hydrogen peroxide H\textsubscript{2}O\textsubscript{2} formation (Figure \ref{scatterplots}b) through the abstraction  reaction  HO\textsubscript{2}+CH\textsubscript{3}OH$\rightarrow$H\textsubscript{2}O\textsubscript{2}+CH\textsubscript{2}OH, extensively investigated by other authors \cite{li2007comprehensive}. At $Z<Z_{st}$, OH recombination reaction  (2OH$\rightarrow$H\textsubscript{2}O\textsubscript{2}) is the main responsible for H\textsubscript{2}O\textsubscript{2} formation in the lean region. At $Z\sim0.35$ there is an intermediate peak (not clearly  visibile due to the very low concentration), still due to the flame quenching at the fiber which locally cools the flame.

\begin{figure}
	\centering
	\subfloat[]
	{\includegraphics[width=.39\textwidth,height=.3\textwidth]{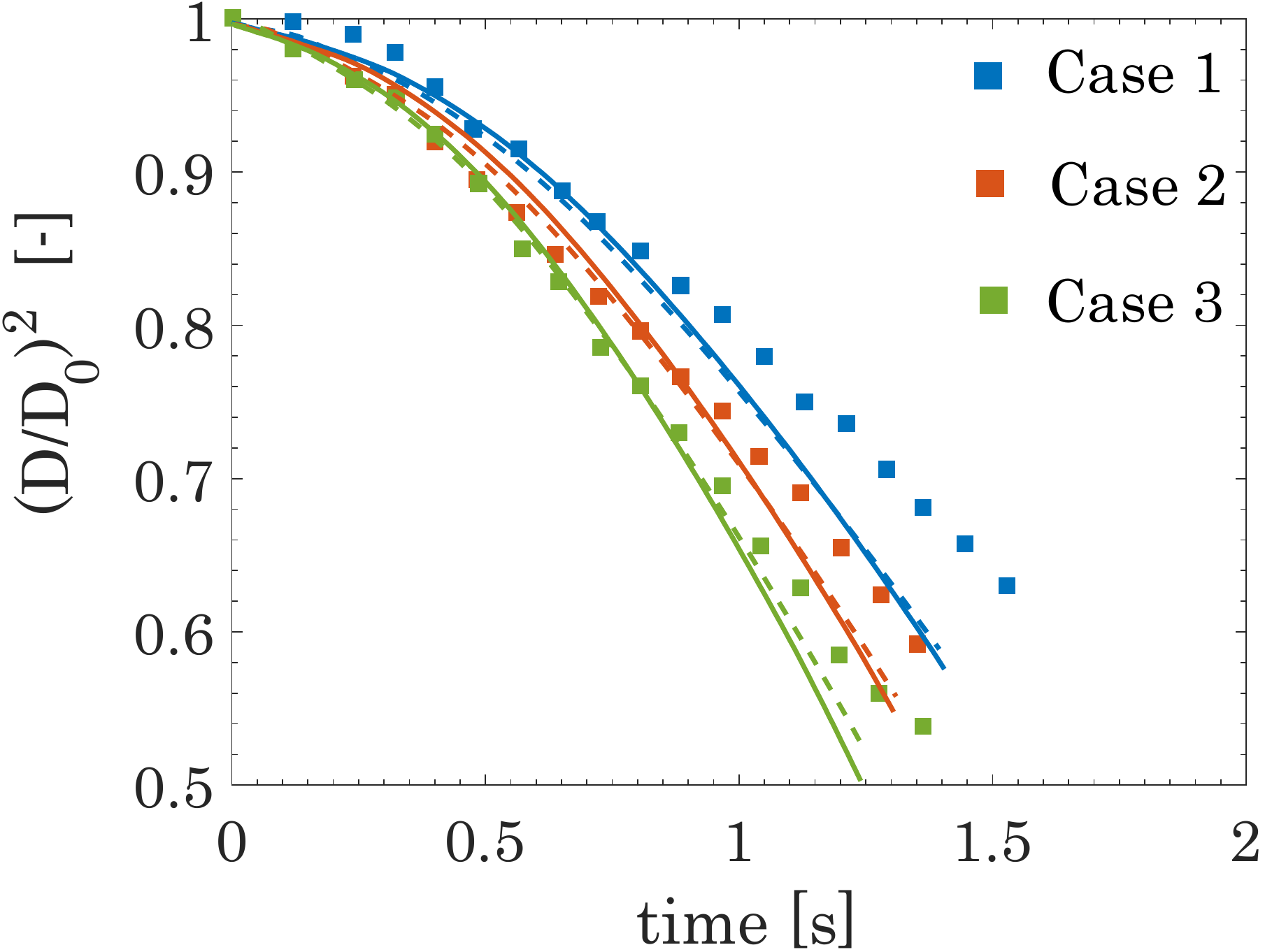}}\quad\quad
	\subfloat[]
	{\includegraphics[width=.39\textwidth,height=.3\textwidth]{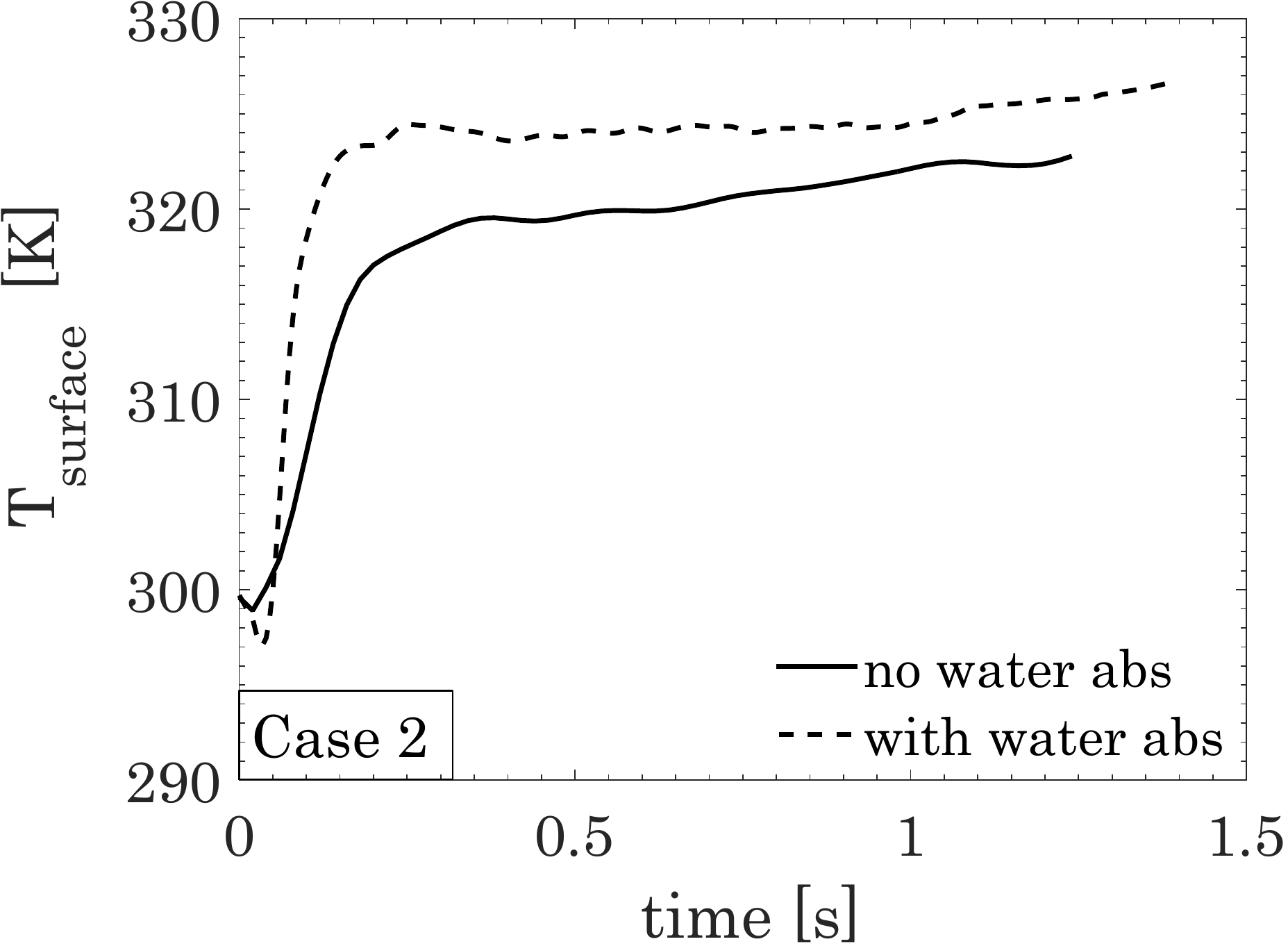}}\\	 
	\subfloat[]
	{\includegraphics[width=.39\textwidth,height=.3\textwidth]{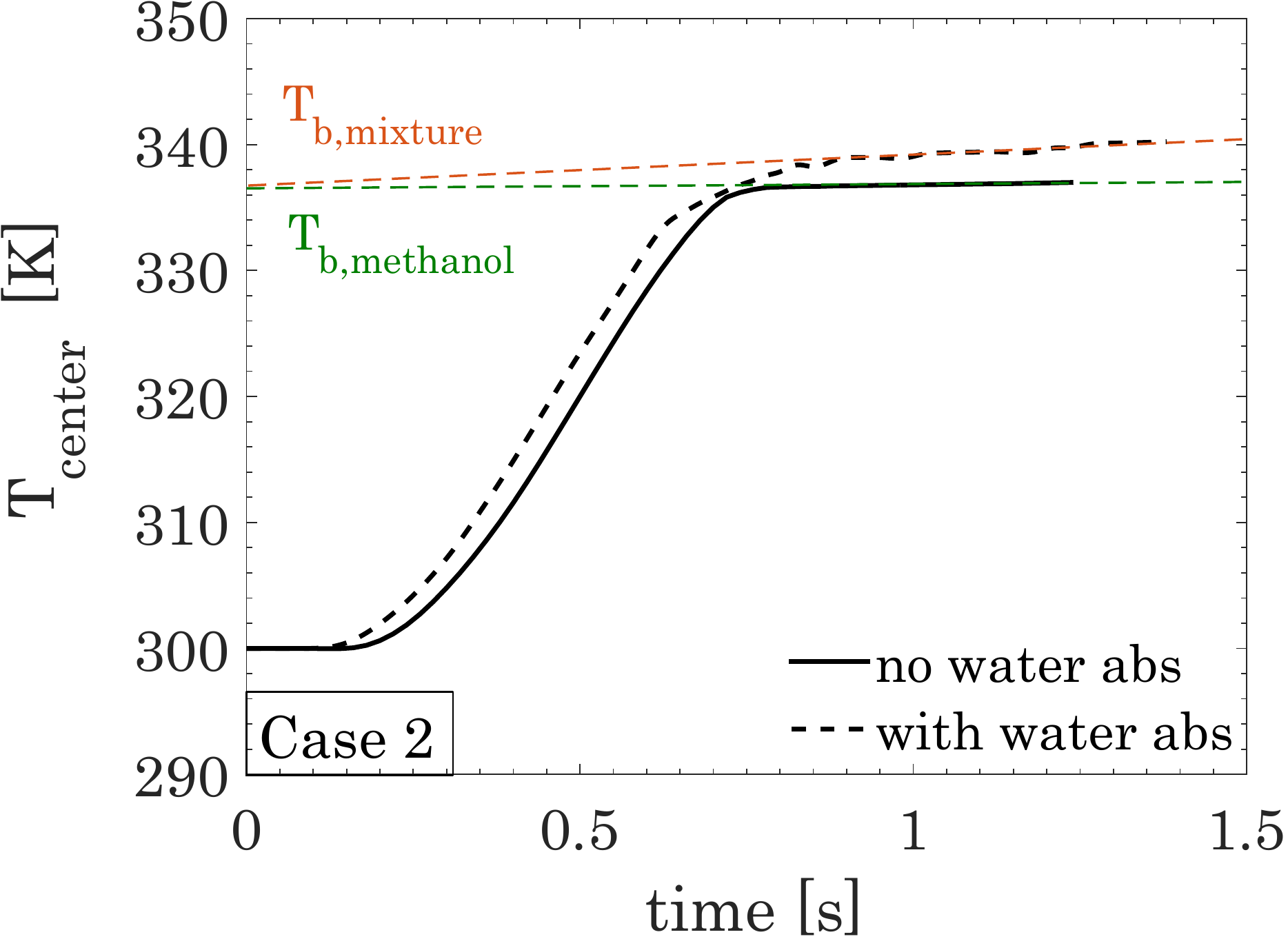}}	 \quad\quad
	\subfloat[]
	{\includegraphics[width=.39\textwidth,height=.3\textwidth]{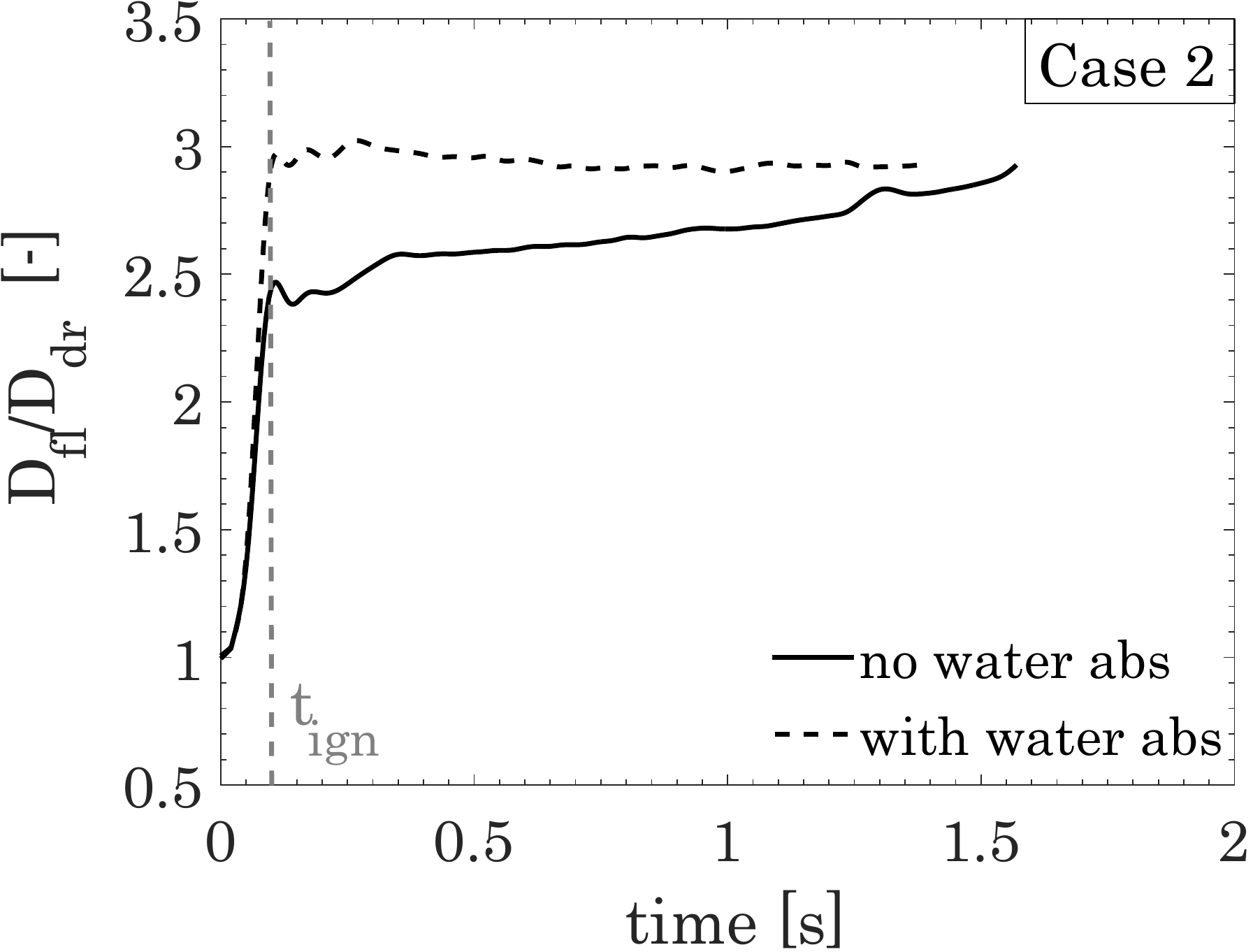}} 
	\caption{Effect of condensed water: $\left(D/D_0\right)^2$ plot of Cases 1, 2, 3 (a), surface temperature profiles (b), internal temperature profiles (c) and standoff ratio $D_{fl}/D_{dr}$ profiles (d) of Case 2.}
	\label{waterAbsorption}
\end{figure}

 The main combustion products are H\textsubscript{2}O and CO\textsubscript{2} (Figure \ref{scatterplots}c), while CO is almost completely oxidized by  OH at the flame front. With respect to the case without fiber, we notice a lower concentration of these species.  It is worth noticing that CO, CO\textsubscript{2} and H\textsubscript{2}O are present in significant amount at the droplet interface ($Z=1$), because of back diffusion from the flame. Differently from CO and CO\textsubscript{2}, water can condense on the droplet surface due to the low interface temperature  ($\sim$ 320 K). This will be analyzed in the next section.

\subsection{Condensation of water}
The water present among the combustion products in the gas-phase is miscible with methanol and it can condense on the droplet surface. If the liquid temperature is sufficiently high, water starts vaporizing as well and a two-component thermodynamics is established. This has been widely investigated in recent numerical and experimental works regarding alcohol combustion \cite{lee1992experimental, farouk2012extinction}. From the modeling point of view it is only necessary to include the liquid species equation (Equation \ref{speciesequationMulticomponent}) and the correct evaporation rate for multicomponent mixtures (Table \ref{evaporationCases}).  As already reported, the boiling temperature $T_b$ is needed to trigger the boiling sub-model: in a two-component mixture (water and methanol) $T_b$ is not fixed with pressure, but it also depends on the local composition and it has to be calculated in every point in the liquid phase at every time step.\par In order to highlight the effect of the absorbed water on the droplet combustion, we simulated the three cases in Table \ref{tableExperimentalCases} including the water condensation flux. The results are reported in Figure \ref{waterAbsorption}. Referring to Figure \ref{waterAbsorption}a, we can identify two main  vaporization regimes of the methanol-water mixture. During the first period the droplet is consumed faster when water absorption is accounted for: this happens because of the condensation heat released on the surface, which increases the temperature ($\sim$4-5 K). Since methanol is (initially) the main component of the two-phase mixture, it vaporizes faster \cite{lee1992experimental}. This can be clearly seen in Figure \ref{waterAbsorption}b, where the surface temperatures for the two cases are reported. Later on, the amount of condensed water becomes significant and the droplet global vaporization rate is retarded (also because of methanol dilution). This effect seems to be more enhanced at high temperatures (Case 3) due to the higher water concentration in the gas-phase. Moreover, the water absorbed at the surface is transported (by diffusion and internal convection) inside the liquid phase, close to the fiber. The vaporization enthalpy of the methanol-water mixture locally increases ($\Delta h_{ev,water}>\Delta h_{ev,meth}$) and the internal boiling flux is diminished (Equation \ref{boilingRate}), further slowing the vaporization rate. At the droplet center the boiling conditions are reached for both cases ($T=T_b$): while $T_b$ is constant for pure methanol vaporization, it constantly increases for the mixture because of water absorption ($T_{b,water}>T_{b,meth}$) which changes the liquid composition (Figure  \ref{waterAbsorption}c). \par
\begin{figure}
	\centering
	\subfloat[]
	{\includegraphics[width=.4\textwidth,height=.3\textwidth]{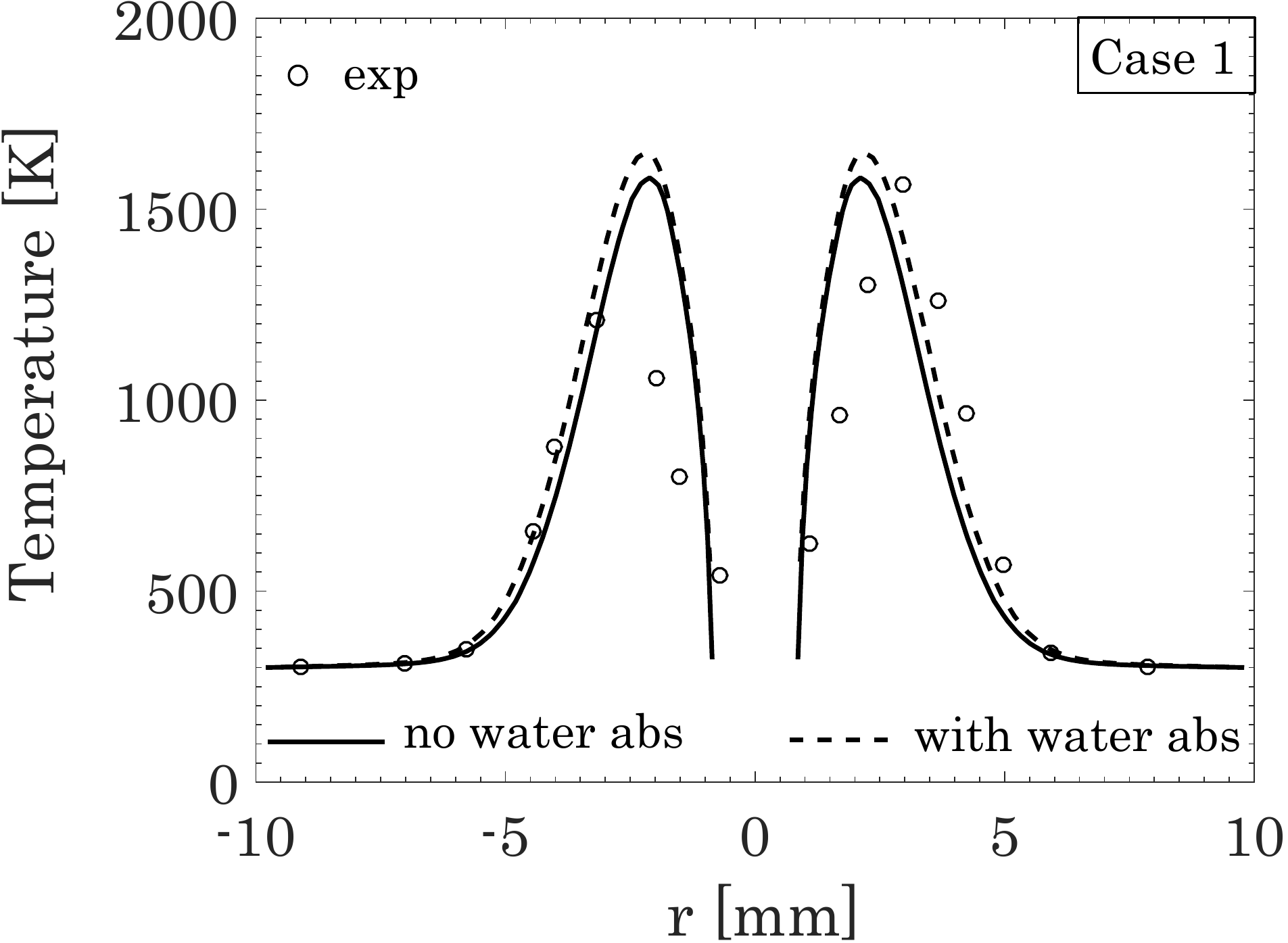}}	 \quad\quad
	\subfloat[] 
	{\includegraphics[width=.4\textwidth,height=.3\textwidth]{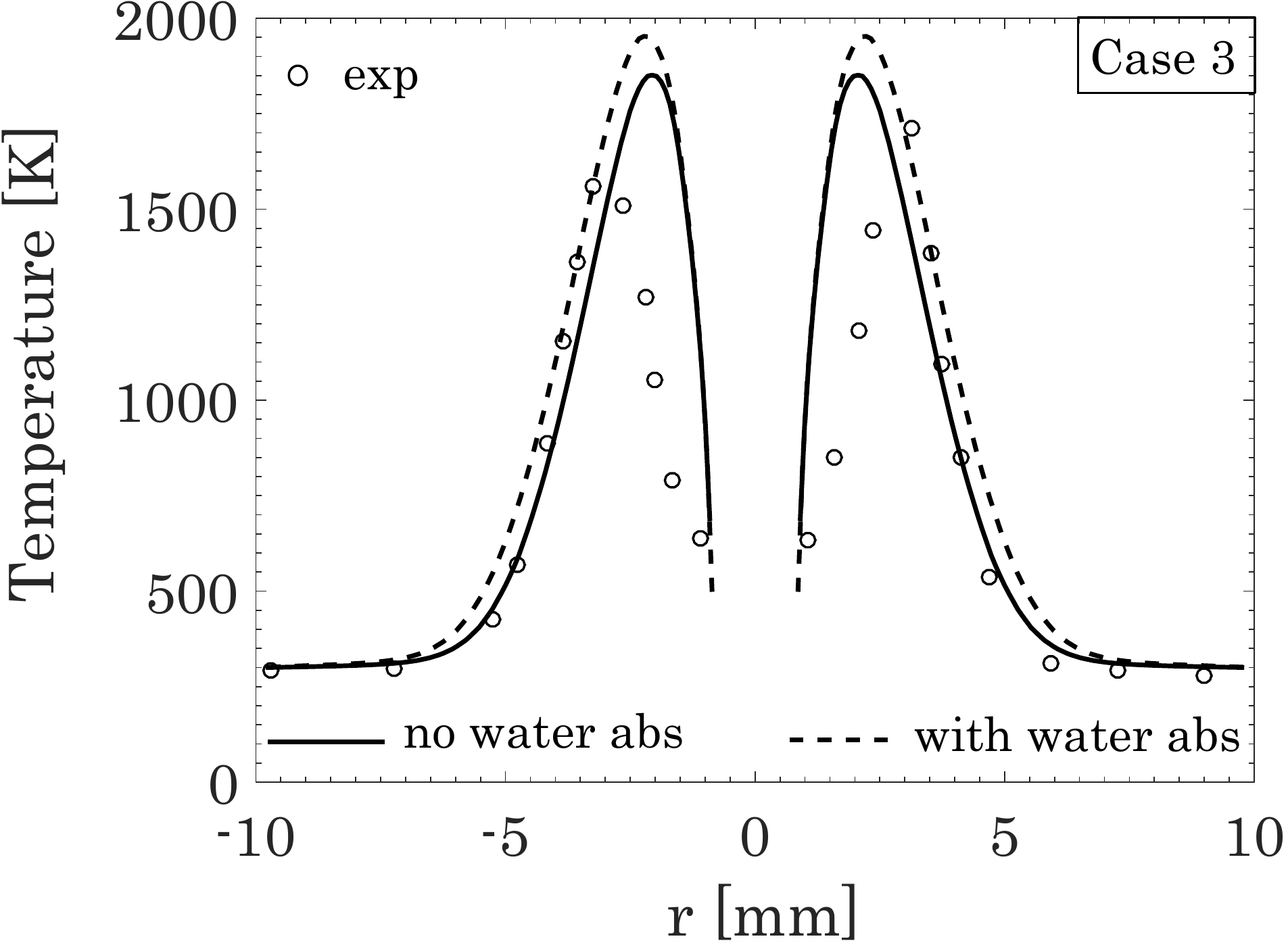}}	 
	\caption{Effect of condensed water: radial temperature profiles for Case 1 (a) and Case 3 (b).}
	\label{radialTwaterAbsorption}
\end{figure}

 It is also worth analyzing the difference in the flame position with respect to the droplet surface (Figure  \ref{waterAbsorption}d). The standoff ratio is higher due to the more intense methanol vaporization, which pushes the flame farther. This is reflected by the comparison of the radial temperature profiles (previously shown in Figure \ref{profilesII}b, c) in Figures \ref{radialTwaterAbsorption}a, b: when water absorption is included, the temperature maximum slightly shifts away from the droplet and improves the agreement with the experimental data. Lee and Law \cite{lee1992experimental} also report the existence of a final evaporation stage in which the condensed water starts to vaporize again once reached a significant concentration in the liquid. However, we did not notice this phenomena in our case probably because of the relatively low amount of water absorbed. In particular, we predict less than 10$\%$ (in mass) of water accumulation in the liquid phase at the end of the simulation (for all the cases). We found this effect to be negligible on the diameter decay (Figure \ref{waterAbsorption}a), since the vaporization rate is controlled by the internal boiling (mainly methanol) induced by the fiber.\par 
To conclude, we want to mention the possibility to observe Marangoni effects in this configuration, i.e. internal flows caused by the gradients of surface tension along the interface \cite{hicks2010methanol}. However, differently from microgravity systems, where this effect can significantly impact the combustion properties, in our cases the liquid motion is due to the external gas-phase convection and to the consequent shear-stress at the interface. Therefore, it is reasonable to assume the Marangoni flow to have a reduced impact on our simulation, since it is  typically dominant in zero or very low  Reynolds numbers ($Re<5$) configurations \cite{raghavan2006surface}. It is clear, however, that further investigations are needed: in particular towards the possibility of introducing advanced surface tension sub-models in multiphase CFD codes for droplet combustion in order to shed light on important effects such as Marangoni flows, contact angle effects and wetting phenomena.

\section{Conclusions}
In this paper we presented an interface-resolved numerical simulation of a methanol droplet suspended on a fiber in normal gravity, using a recent experimental work as a reference case. The multiphase VOF-based solver \texttt{DropletSMOKE++} is adopted for the numerical modeling, because of its capability in describing in detail several physical phenomena such as the interface advection, the two-phase velocity field, the phase-change, the non-ideality of multicomponent mixtures, the  combustion chemistry and the thermal interaction with the fiber.  The high level of detail of the simulation allows to investigate interesting physical aspects related to suspended droplet combustion, in particular:

\begin{itemize}
	\item The thermal perturbation of the fiber strongly affects the vaporization rate, conducting heat inside the liquid phase. The droplet interior is subjected to boiling, while the external surface evaporates by diffusion. A partial quenching of the flame occurs close to its surface, accumulating oxidized species;
	\item Higher oxygen concentrations increase the flame temperature,  bring the flame closer to the droplet surface and diminish the droplet lifetime;
    \item The presence of a buoyant flow provides an axisymmetric flame geometry, with a lower standoff ratio and a higher flame temperature if compared to microgravity. Furthermore, the shear stress at the interface induces an internal circulation in the liquid phase, which enhances the internal heat transfer;
    \item  The flame is thinner in the lower part of the droplet (high $\chi$), where we found the peak of most of the radical species. A brief analysis of the distribution of species in the gas phase shed light on the chemical behavior of H and HO\textsubscript{2} radicals and on the quenching effect at the fiber;
    
    \item The water produced in the gas-phase condenses on the droplet surface, with two main effects: (i) an initial increase of the vaporization rate, due to the release of the condensation enthalpy at the droplet surface and (ii) a subsequent delay, because of water accumulation. The effect on the diameter decay is negligible, while the comparison of the radial temperature profiles slightly improves.
\end{itemize}

To conclude, we want to stress the importance of  such fundamental analyses on isolated droplets  to: (i) provide a general numerical support for the intense experimental activity on single droplet vaporization/combustion, including possible ways of improvement in the design of these systems (e.g. minimize the effect of the fiber) and (ii) the relevance for reduced numerical descriptions of sprays (e.g. LES), in particular for the development of accurate and reliable sub-models for mass transfer rates, drag force, Stefan flow, internal circulation etc. to be used in such complex numerical simulations. \par In addition, we want to underline the generality and flexibility of this approach, which allows to easily introduce more complex kinetic mechanisms to investigate additional phenomena such as low temperature chemistry, soot formation and preferential vaporization, as well as the possibility to investigate  multiple and interacting droplets. 

\section*{Acknowledgments}
We acknowledge the CINECA award under the ISCRA initiative, for the availability of high performance computing resources and support (ISCRA-B: HP10BGXWCZ).

\begin{table}
	\centering
	\begin{tabular}{lcccll}
		\toprule
		\multirow{2}*{Variable} \quad\quad\quad & \multicolumn{3}{c}{Refinement levels}  & \quad\quad \multirow{2}*{$\infty$}   &\quad\quad \multirow{2}*{O} \\
		\cmidrule(lr){2-4}
		& n=1 \quad\quad & n=2  \quad\quad& n=3 \\
		\midrule
		T\textsubscript{flame} & 1711.29 \quad\quad & 1760.2 \quad\quad & 1759.95\quad\quad &  \quad\quad 1759.83 & \quad\quad 2.2 \\
		T\textsubscript{surface} & 316.36 \quad\quad & 319.21\quad\quad & 319.24\quad\quad  & \quad\quad 319.25 & \quad\quad 2.3 \\
		CO\textsubscript{max} & 0.109 \quad\quad & 0.118 \quad\quad & 0.117 \quad\quad  & \quad\quad 0.1165  & \quad\quad 1.8 \\
		CO\textsubscript{2,max} & 0.248 \quad\quad & 0.236 \quad\quad & 0.234 \quad\quad &  \quad\quad 0.233  & \quad\quad 1.8 \\
		H\textsubscript{2}O\textsubscript{max} & 0.171 \quad\quad & 0.146 \quad\quad & 0.143 \quad\quad &  \quad\quad 0.142  & \quad\quad 2.3 \\
		OH\textsubscript{max} & 0.00311 \quad\quad & 0.00373 \quad\quad & 0.00364 \quad\quad &  \quad\quad 0.0036  & \quad\quad 1.7 \\
		\bottomrule
	\end{tabular}
	\caption{Grid refinement analysis for Case 2 at three different levels of resolution. The reference value $\infty$ is estimated through Richardson extrapolation \cite{roache1993completed}. The approximate order of convergence O is also reported.}
	\label{meshRefinement}
	
\end{table}

\appendix
\section{Grid refinement analysis}	
The sensitivity of the numerical results with respect to the mesh size is investigated. Case 2 (Table \ref{tableExperimentalCases}) is simulated at three levels of refinement: n=1 (37,000 cells), n=2 (92,000 cells) and n=3 (160,000 cells). The mesh is refined only in the fluid region, maintaining the proportions between the fine region (around the droplet) and the coarser one (outside the droplet).  The analysis is reported in Table \ref{meshRefinement} at time $t=0.5$ s. The values are compared with the continuum value at zero grid spacing, evaluated adopting the Richardson extrapolation \cite{roache1993completed}. The code shows convergence orders between 1.7 and 2.3 for the examined variables. Even though n=3 is the most accurate case, the refinement level n=2 (92,000 cells) has been used in this work, since the computational time is significantly reduced. Adopting this resolution, the average error for the reported variables remains  between $0.01\%$ and $2\%$.


\nomenclature[G]{$\rho$ }{density $\left[\frac{kg}{m^3}\right]$}
\nomenclature[G]{$\gamma$ }{activity coefficient $\left[-\right]$}
\nomenclature[S]{$L$}{liquid}
\nomenclature[S]{$G$}{gas}
\nomenclature[R]{$T$ }{temperature $\left[K\right]$}
\nomenclature[R]{$\textbf{v}$ }{velocity $\left[\frac{m}{s}\right]$}
\nomenclature[R]{$\dot{m}$ }{evaporation flux $\left[\frac{kg}{m^3s}\right]$}
\nomenclature[R]{$p$ }{pressure $\left[Pa\right]$}
\nomenclature[R]{$p_{rgh}$ }{dynamic pressure $\left[Pa\right]$}
\nomenclature[G]{$\alpha$ }{VOF marker function $[-]$}
\nomenclature[R]{$k$ }{thermal conductivity $\left[\frac{W}{mK}\right]$}
\nomenclature[R]{$\textbf{j}$ }{diffusion flux $\left[\frac{kg}{m^2s}\right]$}
\nomenclature[S]{$i$ }{$i$-th species}
\nomenclature[S]{$j$ }{$j$-th reaction}
\nomenclature[R]{$C_p$ }{constant pressure specific heat  $\left[\frac{J}{kgK}\right]$}
\nomenclature[G]{$\Delta h_{ev}$ }{evaporation enthalpy $\left[\frac{J}{kg}\right]$}
\nomenclature[G]{$\omega$}{mass fraction $\left[-\right]$}
\nomenclature[R]{$t$}{time $\left[s\right]$}
\nomenclature[R]{$M_w$}{molecular weight $\left[\frac{kg}{mol}\right]$}
\nomenclature[R]{$\mathcal{D}$}{mass diffusion coefficient $\left[\frac{m^2}{s}\right]$}
\nomenclature[R]{$x$}{mole fraction $\left[-\right]$}
\nomenclature[R]{$\dot{q}$}{heat flux per unit volume $\left[\frac{W}{m^3}\right]$}
\nomenclature[R]{$R$}{reaction rate $\left[\frac{kg}{m^3s}\right]$}
\nomenclature[G]{$\Delta H_R$}{enthalpy of reaction $\left[J/kg\right]$}
\nomenclature[G]{$\nu$}{stoichiometric coefficient $\left[-\right]$}
\nomenclature[R]{$p^0$}{vapor pressure $\left[Pa\right]$}
\nomenclature[G]{$\phi$}{pure gas-phase fugacity coefficient $\left[-\right]$}
\nomenclature[G]{$\hat{\phi}$}{mixture gas-phase fugacity coefficient $\left[-\right]$}
\nomenclature[R]{$v$}{molar volume $\left[\frac{m^3}{mol}\right]$}
\nomenclature[R]{$\textbf{f}$}{force per unit volume $\left[\frac{N}{m^3}\right]$}
\nomenclature[R]{$\textbf{x}$}{position vector $\left[m\right]$}
\nomenclature[G]{$\kappa$}{curvature $\left[\frac{1}{m}\right]$}
\nomenclature[G]{$\delta_s$}{Dirac delta $\left[\frac{1}{m}\right]$}
\nomenclature[R]{$r$}{radius $\left[m\right]$}
\nomenclature[G]{$\mu$}{dynamic viscosity $\left[\frac{kg}{ms}\right]$}
\nomenclature[R]{$D$}{diameter $\left[m\right]$}
\nomenclature[R]{$Ns$}{Number of species $\left[-\right]$}
\nomenclature[R]{$L$}{base radius of the mesh $\left[m\right]$}
\nomenclature[R]{$H$}{height of the mesh $\left[m\right]$}
\nomenclature[R]{$Z$}{mixture fraction $\left[-\right]$}
\nomenclature[G]{$\chi$}{scalar dissipation rate $\left[\frac{1}{s}\right]$}
\nomenclature[R]{$a_p$}{absorption coefficient $\left[-\right]$}
\nomenclature[G]{$\psi$}{generic variable}
\nomenclature[S]{$s$}{solid}
\nomenclature[S]{$f$}{fiber}
\nomenclature[S]{$b$}{boiling}
\nomenclature[S]{$rad$}{radiative}
\nomenclature[S]{$0$}{initial, reference}
\nomenclature[G]{$\sigma$}{surface tension $\left[\frac{N}{m}\right]$}
\nomenclature[S]{$env$}{ambient}
\nomenclature[A]{VOF}{Volume Of Fluid}

\bibliography{bibliografia}
\bibliographystyle{elsarticle-PROCI}

\end{document}